\documentclass{emulateapj}
\usepackage{longtable}
\usepackage{amsmath}

\def\kms{\ifmmode{\rm km\thinspace s^{-1}}\else km\thinspace s$^{-1}$\fi}
\def\vstar{V530\,Ori}

\shortauthors{Torres et al.}
\shorttitle{\vstar}

\begin{document}

\submitted{Accepted for publication in the Astrophysical Journal}

\title{The G+M eclipsing binary V530\,Orionis: A stringent test of
  magnetic stellar evolution models for low-mass stars}

\author{
Guillermo Torres\altaffilmark{1},
Claud H.\ Sandberg Lacy\altaffilmark{2},
Kre\v{s}imir Pavlovski\altaffilmark{3},
Gregory A.\ Feiden\altaffilmark{4}, 
Jeffrey A.\ Sabby\altaffilmark{5,6},
Hans Bruntt\altaffilmark{7,8}, and
Jens Viggo Clausen\altaffilmark{9,10}
}

\altaffiltext{1}{Harvard-Smithsonian Center for Astrophysics, 60
  Garden St., Cambridge, MA 02138, USA; e-mail:
  gtorres@cfa.harvard.edu}

\altaffiltext{2}{Department of Physics, University of Arkansas,
  Fayetteville, AR 72701, USA}

\altaffiltext{3}{Department of Physics, Faculty of Science, University
  of Zagreb, Bijeni\u{c}ka cesta 32, 10000 Zagreb, Croatia}

\altaffiltext{4}{Department of Physics \& Astronomy, Uppsala
  University, Box 516, SE-751 20 Uppsala, Sweden}

\altaffiltext{5}{Physics Department, Southern Illinois University
  Edwardsville, Edwardsville, IL 62026, USA}

\altaffiltext{6}{Visiting Astronomer, Kitt Peak National Observatory,
  National Optical Astronomy Observatories, operated by the
  Association of Universities for Research in Astronomy, Inc., under a
  cooperative agreement with the National Science Foundation.}

\altaffiltext{7}{Stellar Astrophysics Centre, Department of Physics
  and Astronomy, Aarhus University, Ny Munkegade 120, 8000 Aarhus C,
  Denmark}

\altaffiltext{8}{Aarhus Katedralskole, Skolegyde 1, 8000 Aarhus C,
  Denmark}

\altaffiltext{9}{Niels Bohr Institute, Copenhagen University, Juliane
  Maries Vej 30, DK-2100 Copenhagen {\O}, Denmark}
\altaffiltext{10}{Deceased 2011 June 5.}

\begin{abstract}

We report extensive photometric and spectroscopic observations of the
6.1-day period, G+M-type detached double-lined eclipsing binary
\vstar, an important new benchmark system for testing stellar
evolution models for low-mass stars. We determine accurate masses and
radii for the components with errors of 0.7\% and 1.3\%, as follows:
$M_{\rm A} = 1.0038\pm0.0066$\,$M_{\sun}$, $M_{\rm B} =
0.5955\pm0.0022$\,$M_{\sun}$, $R_{\rm A} = 0.980\pm0.013$\,$R_{\sun}$,
and $R_{\rm B} = 0.5873\pm0.0067$\,$R_{\sun}$.  The effective
temperatures are $5890\pm100$\,K (\ion{G1}{5}) and $3880\pm120$\,K
(\ion{M1}{5}), respectively. A detailed chemical analysis probing more
than 20 elements in the primary spectrum shows the system to have a
slightly subsolar abundance, with ${\rm [Fe/H]} = -0.12\pm0.08$. A
comparison with theory reveals that standard models underpredict the
radius and overpredict the temperature of the secondary, as has been
found previously for other M dwarfs. On the other hand, models from
the Dartmouth series incorporating magnetic fields are able to match
the observations of the secondary star at the same age as the primary
($\sim$3\,Gyr) with a surface field strength of $2.1\pm0.4$\,kG when
using a rotational dynamo prescription, or $1.3\pm0.4$\,kG with a
turbulent dynamo approach, not far from our empirical estimate for
this star of $0.83\pm0.65$\,kG. The observations are most consistent
with magnetic fields playing only a small role in changing the global
properties of the primary. The \vstar\ system thus provides an
important demonstration that recent advances in modeling appear to be
on the right track to explain the long-standing problem of radius
inflation and temperature suppression in low-mass stars.

\end{abstract}


%%%%%%%%%%%%%%%%%%%%%%%%%%%%%%%%%%%%%%%%%%%%%%%%%%%%%%%%%%%%%%%%%%%%%%%%%%%
\section{Introduction}
\label{sec:introduction}
%%%%%%%%%%%%%%%%%%%%%%%%%%%%%%%%%%%%%%%%%%%%%%%%%%%%%%%%%%%%%%%%%%%%%%%%%%%

The discovery of \vstar\ (HD\,294598, BD$-$03~1283,
2MASS\,J06043380$-$0311513) as an eclipsing binary was made by
\cite{Strohmeier:59}, who established an orbital period for the system
of 6.110792 days. The depth reported for the primary eclipse was about
0.7 mag, but no secondary eclipse was seen in these early photographic
measurements. The primary star is of solar type. The object has
received little attention following the discovery, other than the
occasional measurement of times of primary eclipse, which was the only
eclipse detected until recently. It was claimed by \cite{Sahade:63} to
be a possible member of the Collinder\,70 cluster, a proposal that
appears to have since been dismissed. Faint spectral lines of the
secondary with about the same width as those of the primary were first
detected in 1985 by \cite{Lacy:90}, but remained elusive in subsequent
high-resolution observations \citep[see, e.g.][]{Popper:96}.
Similarly, no signs of the secondary eclipse could be seen in more
recent photometric monitoring, implying either a very faint and cool
companion, or possibly an eccentric orbit and a special orientation
such that no secondary eclipses occur.

This motivated us to begin our own program of spectroscopic
observation in 1996. Our interest in the system was piqued when we
were able to derive the first single-lined spectroscopic orbit, which
is indeed eccentric but only slightly so, and to predict the exact
location of the secondary eclipse, which we were then successful in
detecting with more targeted photometric observations. The depth in
$V$ is less than 3\%.  Continued analysis has enabled us to also
measure radial velocities for the secondary, and to fully characterize
the binary.

The confirmed presence of a late-type star in \vstar\ makes it a rare
example of a system containing a solar-type primary that is easy to
study and provides access to other key properties of the binary, and
at the same time a late-type secondary that is very faint but still
measurable. As such, \vstar\ is potentially very useful for testing
models of stellar evolution if accurate properties for the stars can
be derived, by virtue of the greater leverage afforded by a mass ratio
significantly different from unity.  Previous measurements for M
dwarfs have shown rather serious disagreements with models in the
sense that such stars appear larger and cooler than predicted by
theory \citep[e.g.,][]{TorresRibas:02, Ribas:03, LopezMorales:05,
  Torres:13}. This is now widely believed to be related to stellar
activity \citep[magnetic inhibition of convection, and/or star
  spots;][]{Mullan:01, Chabrier:07, Feiden:12}, but there are
relatively few systems containing M stars with complete information
available for testing this hypothesis.

Here we provide a full description of our spectroscopic and
photometric observations of \vstar, leading to the first determination
of accurate properties for the stars including the absolute masses and
radii. We report also a detailed chemical analysis of the system based
on the solar-type primary star, bypassing the usual difficulties and
limitations of determining the metallicity of M stars.  We
additionally estimate the surface magnetic field strengths for both
components, an important piece of information permitting a more
meaningful comparison with recent models that incorporate magnetic
fields. Our results provide one of the clearest illustrations that
such models are indeed able to reproduce the measured properties of
low-mass stars.

%%%%%%%%%%%%%%%%%%%%%%%%%%%%%%%%%%%%%%%%%%%%%%%%%%%%%%%%%%%%%%%%%%%%%%%%%%%
\section{Ephemeris}
\label{sec:ephemeris}
%%%%%%%%%%%%%%%%%%%%%%%%%%%%%%%%%%%%%%%%%%%%%%%%%%%%%%%%%%%%%%%%%%%%%%%%%%%

Dates of minimum light for \vstar\ were collected from the literature
and from our own unpublished photometric measurements (see
Table~\ref{tab:tmin}), and were used to establish the ephemeris. The
measurements (34 timings for the primary and 7 for the secondary) span
about 82 years, or $\sim$4900 orbital cycles of the binary.
Uncertainties for the older timings and for some of the more recent
ones have not been published, so we determined them by iterations to
achieve reduced $\chi^2$ values near unity, separately for each type
of measurement ($\sigma = 0.028$, 0.011, and 0.0001 days for the
photographic, visual, and photoelectric/CCD data). We found we also
needed to rescale the published photoelectric/CCD errors by factors of
1.8 and 2.9 for the primary and secondary, respectively. A linear
weighted least-squares fit using the primary and secondary minima
together resulted in
\begin{align*}
&{\rm Min~I~(HJD)} = 2,\!453,\!050.826061(91) + 6.11077840(33) E \\
&{\rm Min~II~(HJD)} = 2,\!453,\!053.6623(16) + 6.11077840(33) E\,,
\end{align*}
which we have used in the analysis that follows. Uncertainties are
indicated in parentheses in units of the last significant digit.

\begin{figure}
\epsscale{1.10}
\plotone{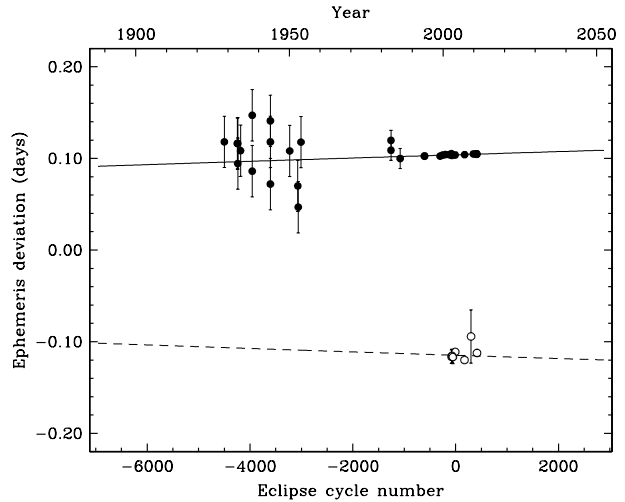}

\figcaption{Ephemeris curve for \vstar\ from the fit described in the
  text. Times of eclipse are shown with filled circles for the primary
  and open circles for the secondary. Eclipse cycles are counted from
  the reference epoch given in the text. The corresponding apsidal
  period is $U = 7800 \pm 22000$ years.\label{fig:apsidal}}

\end{figure}

Secondary eclipses occur at a phase of $0.46414(27)$, clearly showing
that the orbit is eccentric. Some degree of apsidal motion is
therefore expected.  An ephemeris curve \citep{Lacy:92a} was fit to
all the data with the same weighting scheme as above, adopting values
for the eccentricity and inclination angle derived in our
spectroscopic and light curve analyses below, and is illustrated in
Figure~\ref{fig:apsidal}. However, the apsidal period is only poorly
determined from this fit ($U = 7800 \pm 22000$ years).

\begin{deluxetable*}{clcrcccc}
\tablewidth{0pc}
\tablecaption{Times of eclipse for \vstar.\label{tab:tmin}}
\tablehead{
\colhead{} &
\colhead{HJD} &
\colhead{$\sigma$\tablenotemark{a}} &
\colhead{} &
\colhead{} &
\colhead{$(O-C)$} &
\colhead{} &
\colhead{} \\
\colhead{Year} &
\colhead{(2,400,000$+$)} &
\colhead{(days)} &
\colhead{Epoch\tablenotemark{b}} &
\colhead{Ecl\tablenotemark{c}} &
\colhead{(days)} &
\colhead{Type\tablenotemark{d}} &
\colhead{Source\tablenotemark{e}} }
\startdata
  1928.8527 &  25558.456   &   \nodata  &   $-$4499 &  1 &  $+$0.0220  &  PG &  1 \\
  1933.0688 &  27098.370   &   \nodata  &   $-$4247 &  1 &  $+$0.0198  &  PG &  1 \\
  1933.2193 &  27153.345   &   \nodata  &   $-$4238 &  1 &  $-$0.0022  &  PG &  1 \\
  1933.2193 &  27153.367   &   \nodata  &   $-$4238 &  1 &  $+$0.0198  &  PG &  1 \\
  1934.1228 &  27483.341   &   \nodata  &   $-$4184 &  1 &  $+$0.0118  &  PG &  1
\enddata

\tablecomments{This table is available in its entirety in
  machine-readable and Virtual Observatory (VO) forms in the online
  journal. A portion is shown here for guidance regarding its form and
  content.}

\tablenotetext{a}{Timing uncertainties as published, or as measured in
  the case of our own photometric observations. Adopted uncertainties
  for the photographic, visual, and photoelectric/CCD measurements
  with no published errors are 0.028, 0.011, and 0.0001 days,
  respectively. Other errors have been scaled by iterations during the
  ephemeris fit by factors of 1.8 and 2.9 for the primary and
  secondary (see text).}

\tablenotetext{b}{`Epoch' refers to the cycle number counted from the
  reference time of primary eclipse (see text).}

\tablenotetext{c}{`Ecl' is 1 for primary eclipses and 2 for secondary
  eclipses.}

\tablenotetext{d}{`Type' is PG for photographic, V for visual, and PE
  for photoelectric or CCD measurements.}

\tablenotetext{e}{Sources are: (1) \cite{Strohmeier:59}; (2)
  \cite{Isles:88}; (3) \cite{Lacy:94}; (4) \cite{Lacy:99}; (5) This
  paper; (6) \cite{Lacy:02}; (7) \cite{Lacy:04}; (8) \cite{Lacy:07};
  (9) \cite{Nagai:08}; (10) \cite{Diethelm:09} ; (11) \cite{Lacy:11};
  (12) \cite{Diethelm:11}.}

\end{deluxetable*}

%%%%%%%%%%%%%%%%%%%%%%%%%%%%%%%%%%%%%%%%%%%%%%%%%%%%%%%%%%%%%%%%%%%%%%%%%%%
\section{Spectroscopic observations}
\label{sec:spectroscopy}
%%%%%%%%%%%%%%%%%%%%%%%%%%%%%%%%%%%%%%%%%%%%%%%%%%%%%%%%%%%%%%%%%%%%%%%%%%%

\vstar\ was monitored spectroscopically with three different
instruments over a period of more than 17 years. Observations began at
the Harvard-Smithsonian Center for Astrophysics (CfA) in 1996 June
with a Cassegrain-mounted echelle spectrograph \citep[``Digital
  Speedometer'' (DS);][]{Latham:92} attached to the 1.5\,m Tillinghast
reflector at the F.\ L.\ Whipple Observatory (Mount Hopkins,
AZ). Those observations continued through 2009 April. The spectra
consist of a single order 45\,\AA\ wide recorded with an intensified
photon-counting Reticon detector at a central wavelength of 5187\,\AA,
which includes the \ion{Mg}{1}\,b triplet. The resolving power provided by
this setup is $R \approx 35,\!000$. Additional observations were
collected with a nearly identical instrument attached to the
4.5\,m-equivalent Multiple Mirror Telescope (also on Mount Hopkins),
prior to its conversion to a monolithic 6.5\,m telescope. The 74
usable spectra from these instruments have signal-to-noise ratios
(SNRs) ranging from about 10 to 50 per resolution element of
8.5\,\kms. Observations of the dusk and dawn sky were made every night
to monitor the velocity zero point, and to establish small run-to-run
corrections applied to the DS velocities reported below.

We gathered a further 30 spectra of \vstar\ at the Kitt Peak National
Observatory (KPNO) from 1999 March to 2001 January, using the
coud\'e-feed telescope and the coud\'e spectrometer. The spectra cover
the wavelength region 6450--6600\,\AA, and include the H$\alpha$
line. The 250\,$\mu$m slit and OG\,550 filter projected onto
0.186\,\AA\ on the detector. The detector was a Ford $3072\times1024$
pixel CCD (F3KB) with 15\,$\mu$m square pixels. The `A' grating (632
grooves~mm$^{-1}$) was used in the second order with Camera~5 (a
folded Schmidt design). The spectra were flat-fielded and wavelength
calibrated following standard procedures, based on quartz lamp flats
and Th-Ar emission tube spectra. Observations of the standard stars
$\iota$\,Psc or $\beta$\,Vir were taken with the same setup during the
same nights in order to correct for instrumental drifts. The
adjustments assumed constant velocities of $+5.636$\,\kms\ for
$\iota$\,Psc (HD\,222368) and $+4.468$\,\kms\ for $\beta$\,Vir
(HD\,102870), from \cite{Nidever:02}.

Finally, 41 additional observations were obtained at the CfA from 2009
November to 2014 March with the Tillinghast Reflector Echelle
Spectrograph \citep[TRES;][]{Furesz:08} on the 1.5\,m telescope
mentioned earlier. This bench-mounted instrument yields a resolving
power of $R \approx 44,\!000$, and spectra spanning
3860--9100\,\AA\ in 51 orders. The SNRs range from 13 to 121 per
resolution element of 6.8\,\kms. Instrumental drifts for TRES are
below 10 m\,s$^{-1}$ in velocity, which is negligible for our
purposes.

Lines of the very faint secondary star in \vstar\ are not immediately
obvious in any of our spectra, even in the redder ones from KPNO, but
its radial velocities (RVs) can nevertheless be measured accurately
along with those of the primary using the two-dimensional
cross-correlation algorithm TODCOR \citep{Zucker:94}.  Templates for
the DS and TRES spectra were selected from a large library of
calculated spectra based on model atmospheres by R.\ L.\ Kurucz
\citep[see][]{Nordstrom:94, Latham:02} and a line list prepared by
J.\ Morse. These templates cover approximately 300\,\AA\ centered on
the \ion{Mg}{1}\,b region, and include numerous other lines mainly of
Fe, Ca, and Ti. For the KPNO spectra we used a different template
library based on PHOENIX models \citep[see][]{Husser:13}, kindly
computed for us by I.\ Czekala for the wavelength region of
interest. Our synthetic templates are parametrized in terms of the
effective temperature ($T_{\rm eff}$), rotational velocity ($v \sin i$
when seen in projection), surface gravity ($\log g$), and metallicity,
${\rm [Fe/H]}$. The latter two have a minimal impact on the
velocities, so we adopted fixed values of $\log g = 4.5$ and solar
composition for both stars. The optimum template parameters ($T_{\rm
  eff}$ and $v \sin i$) for the primary were determined following
\cite{Torres:02} by running grids of cross-correlations seeking the
best template match as measured by the mean cross-correlation
coefficient averaged over all exposures. This was done separately for
the three sets of spectra, with very consistent results. We obtained
$T_{\rm eff} = 6000$\,K and $v \sin i = 10$\,\kms.  The faintness of
the secondary, which has a flux some 40 times smaller than that of the
primary, prevents us from determining its template parameters in a
similar way. Instead we relied on the temperature difference inferred
from our light curve solutions in Sect.~\ref{sec:lightcurves}, and we
assumed the star is rotating synchronously. The latter is a reasonable
assumption, as the timescale for synchronization of the secondary
\citep[$\sim$10$^7$\,yr; see, e.g.,][]{Hilditch:01} is much shorter than
the $\sim$3\,Gyr age we estimate for the system later in
Sect.~\ref{sec:models}. With these constraints the template parameters
for the secondary were $T_{\rm eff} = 4000$\,K and $v \sin i =
6$\,\kms.

The final heliocentric velocities from the TRES spectra are the
average of the measurements from the three echelle orders covered by
the templates, and are listed in Table~\ref{tab:RVs}. Typical
uncertainties are 0.05\,\kms\ for the primary (star A) and
1.6\,\kms\ for the faint secondary (star B). Experience has shown that
the very narrow wavelength range of the DS spectra (45\,\AA) can
sometimes lead to systematic errors in the RVs due to residual line
blending as well as lines shifting in and out of the spectral window
as a function of orbital phase \citep[see][]{Latham:96}. We
investigated this by means of numerical simulations for each spectrum,
and found the effect to be significant (shifts of up to 7\,\kms\ for
the secondary, but only 0.02\,\kms\ for the primary). We therefore
applied corrections to the individual velocities in the same way as
done in previous studies with similar spectroscopic material
\citep[e.g.,][]{Torres:97, Lacy:10} in order to remove the bias. These
adjustments increase the minimum masses by about 4\% for the primary
star and 2\% for the secondary. The final DS velocities with
corrections included are given also in Table~\ref{tab:RVs}. They have
typical uncertainties of 0.5\,\kms\ and 6.7\,\kms\ for the primary and
secondary, respectively.  RVs from the KPNO observations are based on
the entire wavelength range of those spectra except for the broad
H$\alpha$ line, which was masked out. Those measurements (two being
excluded here for giving very large residuals from the orbit described
in the next section) are presented with the others in
Table~\ref{tab:RVs}. Their uncertainties are typically 0.4\,\kms\ for
the primary and 5.4\,\kms\ for the secondary.

\begin{deluxetable*}{lcrrrrc}
\tablewidth{0pc}
\tablecaption{Heliocentric radial velocity measurements of \vstar.
 \label{tab:RVs}}
\tablehead{
\colhead{HJD} &
\colhead{Orbital} &
\colhead{$RV_{\rm A}$} &
\colhead{$RV_{\rm B}$} &
\colhead{$(O-C)_{\rm A}$} &
\colhead{$(O-C)_{\rm B}$} &
\colhead{} \\
\colhead{(2,400,000$+$)} &
\colhead{phase} &
\colhead{(\kms)} &
\colhead{(\kms)} &
\colhead{(\kms)} &
\colhead{(\kms)} &
\colhead{Instrument}}
\startdata
    50407.0052  &  0.3513 &   $-$68.18  &   28.42 &    $-$0.41 &   $+$5.33   & DS \\
    50412.8283  &  0.3042 &   $-$78.55  &   43.26 &    $-$0.38 &   $+$2.63   & DS \\
    50441.8571  &  0.0546 &   $-$57.79  &    1.31 &    $-$1.05 &   $-$3.19   & DS \\
    50448.7155  &  0.1770 &   $-$86.70  &   58.24 &    $-$0.65 &   $+$4.33   & DS \\
    50474.7700  &  0.4406 &   $-$43.10  &  $-$17.49 &    $+$0.38 &   $+$0.35   & DS 
\enddata
\tablecomments{This table is available in its entirety in
  machine-readable and Virtual Observatory (VO) forms in the online
  journal. A portion is shown here for guidance regarding its form and
  content.}
\end{deluxetable*}

Our TODCOR analyses also provided an estimate of the light ratio
between the primary and secondary at the mean wavelength of our
spectra \citep[see][]{Zucker:94}. For the DS observations we obtained
$\ell_{\rm B}/\ell_{\rm A} = 0.014 \pm 0.002$ in the \ion{Mg}{1}\,b
region, corresponding to a magnitude difference $\Delta m = 4.6$. The
TRES spectra yielded a similar value of $0.013 \pm 0.002$ for the
average of the three orders used to measure RVs, centered also on the
\ion{Mg}{1}\,b region. As expected from the spectral types, the
secondary appears brighter at the redder wavelengths of the KPNO
spectra, and the light ratio obtained there is $0.042 \pm 0.003$ at a
mean wavelength of 6410\,\AA.

Our TRES spectra display moderately strong emission cores in the
\ion{Ca}{2} H and K lines, which is indicative of stellar
activity. Measurement of the radial velocity of the emission cores
shows that they follow the center of mass of the primary, and are thus
associated with that star. Further evidence of activity is presented
below.

%%%%%%%%%%%%%%%%%%%%%%%%%%%%%%%%%%%%%%%%%%%%%%%%%%%%%%%%%%%%%%%%%%%%%%%%%%%
\subsection{Spectroscopic orbital solution}
\label{sec:orbit}
%%%%%%%%%%%%%%%%%%%%%%%%%%%%%%%%%%%%%%%%%%%%%%%%%%%%%%%%%%%%%%%%%%%%%%%%%%%

Separate spectroscopic orbital solutions using the three velocity data
sets were carried out to check for potential systematic differences,
with the ephemeris held fixed at the values in
Sect.~\ref{sec:ephemeris}.  The results shown in
Table~\ref{tab:orbits} indicate fairly good agreement considering the
faintness of the secondary and the difficulty in measuring its
velocity. Our adopted solution combining all of the RVs is given in
the last column, where we have allowed for arbitrary offsets between
the DS and KPNO velocities relative to those measured with TRES, which
are non-negligible in both cases. The TRES velocities dominate because
of their considerably smaller uncertainties; the rms residuals
($\sigma_{\rm A}$ and $\sigma_{\rm B}$) are listed at the bottom of
the table along with other quantities of interest. We find the orbit
to be slightly eccentric ($e = 0.08802 \pm 0.00023$), consistent with
predictions from theory for this system indicating a timescale for
tidal circularization of $\sim$18\,Gyr \citep[e.g.,][]{Hilditch:01}.

\begin{deluxetable*}{lcccc}
\tablewidth{0pc}
\tablecaption{Spectroscopic orbital solutions for \vstar.\label{tab:orbits}}
\tablehead{
\colhead{\hfil~~~~~~~~~~~~~~~~~Parameter~~~~~~~~~~~~~~~~~} & 
\colhead{TRES} &
\colhead{DS} &
\colhead{KPNO} &
\colhead{Combined}
}
\startdata
$P$ (days)\tablenotemark{a}\dotfill                       &         6.11077840 (fixed)          &   6.11077840 (fixed)              &   6.11077840 (fixed)              &   6.11077840 (fixed)               \\
$\gamma$ (\kms)\dotfill                                   & $-$33.529~$\pm$~0.011\phn\phs       &  $-$33.901~$\pm$~0.070\phn\phs    & $-$32.931~$\pm$~0.079\phn\phs     & \phm{\tablenotemark{b}}$-$33.525~$\pm$~0.011\tablenotemark{b}\phn\phs      \\
$K_{\rm A}$ (\kms)\dotfill                                &  50.9057~$\pm$~0.0083\phn           &   50.986~$\pm$~0.060\phn          &     50.96~$\pm$~0.10\phn          &   50.9075~$\pm$~0.0080\phn         \\
$K_{\rm B}$ (\kms)\dotfill                                &   85.73~$\pm$~0.27\phn              &      87.12~$\pm$~0.85\phn         &      84.8~$\pm$~1.4\phn           &    85.81~$\pm$~0.25\phn            \\
$e$\dotfill                                               &    0.08791~$\pm$~0.00024            &    0.0895~$\pm$~0.0012            &     0.0903~$\pm$~0.0019           &    0.08802~$\pm$~0.00023           \\
$\omega_{\rm A}$ (deg)\dotfill                            &     129.33~$\pm$~0.17\phn\phn       &      129.2~$\pm$~1.1\phn\phn      &      129.5~$\pm$~1.0\phn\phn      &     129.35~$\pm$~0.16\phn\phn      \\
$T$ (HJD$-$2,400,000)\tablenotemark{a}\dotfill            &  53050.826061 (fixed)               &  53050.826061 (fixed)             &  53050.826061 (fixed)             &  53050.826061 (fixed)              \\
$\Delta RV$(TRES$-$DS) (\kms)\dotfill                     &         \nodata                     &          \nodata                  &           \nodata                 &   $+$0.413~$\pm$~0.055\phs         \\
$\Delta RV$(TRES$-$KPNO) (\kms)\dotfill                   &         \nodata                     &          \nodata                  &          \nodata                  &  $-$0.596~$\pm$~0.080\phs          \\[-0.5ex]
                                                                                       
\cutinhead{Derived quantities}                                                                                                                                                                               
$M_{\rm A}\sin^3 i$ (M$_{\sun}$)\dotfill                  &    1.0016~$\pm$~0.0071              &     1.040~$\pm$~0.023             &     0.978~$\pm$~0.035             &   1.0038~$\pm$~0.0066              \\
$M_{\rm B}\sin^3 i$ (M$_{\sun}$)\dotfill                  &    0.5948~$\pm$~0.0024              &    0.6084~$\pm$~0.0076            &     0.588~$\pm$~0.012             &    0.5955~$\pm$~0.0022             \\
$q\equiv M_{\rm B}/M_{\rm A}$\dotfill                     &     0.5938~$\pm$~0.0019             &     0.5852~$\pm$~0.0058           &      0.6009~$\pm$~0.0097          &      0.5932~$\pm$~0.0017           \\
$a_{\rm A}\sin i$ ($10^6$~km)\dotfill                     &      4.26101~$\pm$~0.00069          &      4.2671~$\pm$~0.0051          &       4.2649~$\pm$~0.0087         &      4.26112~$\pm$~0.00067         \\
$a_{\rm B}\sin i$ ($10^6$~km)\dotfill                     &     7.176~$\pm$~0.023               &     7.291~$\pm$~0.071             &      7.10~$\pm$~0.11              &     7.183~$\pm$~0.021              \\
$a \sin i$ (R$_{\sun}$)\dotfill                           &     16.440~$\pm$~0.033\phn          &     16.62~$\pm$~0.10\phn          &      16.33~$\pm$~0.16\phn         &    16.450~$\pm$~0.030\phn          \\[-0.5ex]
                                                                                       
\cutinhead{Other quantities pertaining to the fit}                                                                                                                                                           
$N_{\rm A}$~,~$N_{\rm B}$, TRES\dotfill                   &          41~,~41                    &          \nodata                  &         \nodata                   &          41~,~41                   \\
$N_{\rm A}$~,~$N_{\rm B}$, DS\dotfill                     &         \nodata                     &          74~,~74                  &          \nodata                  &           74~,~74                  \\
$N_{\rm A}$~,~$N_{\rm B}$, KPNO\dotfill                   &         \nodata                     &          \nodata                  &           28~,~28                 &          28~,~28                   \\
Time span (days)\dotfill                                  &          1585.8                     &           4521.6                  &            663.2                  &          6323.7                    \\
$\sigma_{\rm A}$~,~$\sigma_{\rm B}$, TRES (\kms)\dotfill  &        0.049~,~1.66                 &          \nodata                  &         \nodata                   &       0.048~,~1.63                 \\
$\sigma_{\rm A}$~,~$\sigma_{\rm B}$, DS (\kms)\dotfill    &         \nodata                     &        0.46~,~6.65                &           \nodata                 &         0.47~,~6.70                \\
$\sigma_{\rm A}$~,~$\sigma_{\rm B}$, KPNO (\kms)\dotfill  &         \nodata                     &          \nodata                  &         0.42~,~5.47               &         0.42~,~5.37              
\enddata
\tablenotetext{a}{Period and time of primary eclipse from Sect.~\ref{sec:ephemeris}.}
\tablenotetext{b}{Center-of-mass velocity on the reference system of the TRES instrument.}
\end{deluxetable*}

\begin{figure}
\epsscale{1.10}
\plotone{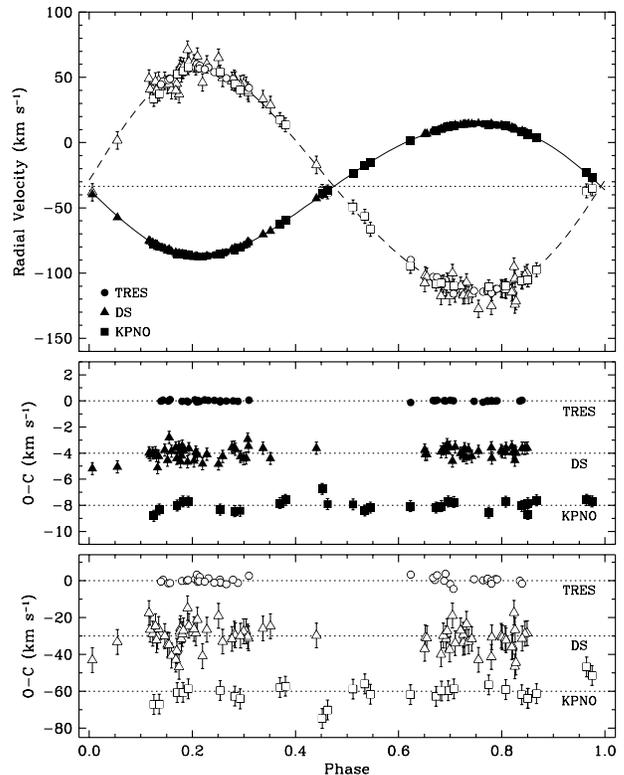}

\figcaption{\emph{Top:} Radial velocities for \vstar\ and our model
  from the combined solution of Table~\ref{tab:orbits} (solid line for
  the primary, dashed for the secondary). The dotted line marks the
  center-of-mass velocity of the system, and phase 0.0 corresponds to
  primary eclipse. Measurements from different data sets are
  represented with different symbols, as labeled. \emph{Middle:}
  Velocity residuals ($O-C$) for the primary star, shown separately
  for each data set. The DS and KPNO residuals are displaced
  vertically for clarity. \emph{Bottom:} Same as middle panel, for the
  secondary.\label{fig:orbit}}

\end{figure}

A graphical representation of our fit appears in
Figure~\ref{fig:orbit} together with the observations and the RV
residuals, the latter shown separately for each data set.

%%%%%%%%%%%%%%%%%%%%%%%%%%%%%%%%%%%%%%%%%%%%%%%%%%%%%%%%%%%%%%%%%%%%%%%%%%%
\subsection{Spectral disentangling}
\label{sec:disentangling}
%%%%%%%%%%%%%%%%%%%%%%%%%%%%%%%%%%%%%%%%%%%%%%%%%%%%%%%%%%%%%%%%%%%%%%%%%%%

Although a number of eclipsing binaries containing M stars have been
studied in the past, in very few cases is the metallicity of the
system known because of the difficulty of analyzing the spectra of
late-type stars, which are dominated by strong molecular features. In
\vstar\ the primary is a solar-type star, for which an abundance
analysis would be straightforward except for the fact that its
spectrum is contaminated at some level by the secondary. To remove
this effect we have subjected our observations to spectral
disentangling \citep{Bagnuolo:91, Simon:94, Hadrava:95}, by which we
are able to reconstruct the spectra of the individual components for
further analysis. \cite{Pavlovski:05} and others have shown that
disentangled spectra can yield reliable abundances \citep[see
  also][]{Pavlovski:10, Pavlovski:12}.

The application of the technique to \vstar\ pushes it to the limit
because of the extreme faintness of the secondary (2.5\% fractional
light in $V$, and even less toward the blue) and the modest SNRs of
our spectra. Some previous studies have succeeded in similar
situations with light ratios of $\sim$5\%
\citep[e.g.,][]{Pavlovskietal:09, Lehmann:13, Tkachenko:14} and even
1.5--2\% \citep{Holmgren:99, Pavlovskietal:10, Mayer:13}, but with
spectra of considerably higher SNR than ours.

We performed disentangling separately for each of our three data sets
(TRES, DS, KPNO) because of their different spectral resolutions and
wavelength coverage, discarding a few spectra with low SNR. We used
the program {\sc FDBinary} \citep{Ilijic:04}, which implements
disentangling in the Fourier domain \citep{Hadrava:95}. For the DS and
KPNO observations we disentangled the entire spectral range available,
and for TRES we restricted ourselves to the interval
4475--6760\,\AA\ to avoid regions with lower flux or telluric
contamination. Special care was taken to select spectral stretches
with both ends in the continuum, as required by the algorithm. Given
the the rich line spectrum the wavelength regions we disentangled
differ in length from 30\,\AA\ to 150\,\AA. Renormalization of the
disentangled spectra \citep[see][]{Pavlovski:05, Lehmann:13} was
performed using the measured light ratios reported earlier from our
spectroscopic analysis as well as those below from our light curve
fits, interpolating or extrapolating linearly as needed.

\begin{figure}
\epsscale{1.15}
\plotone{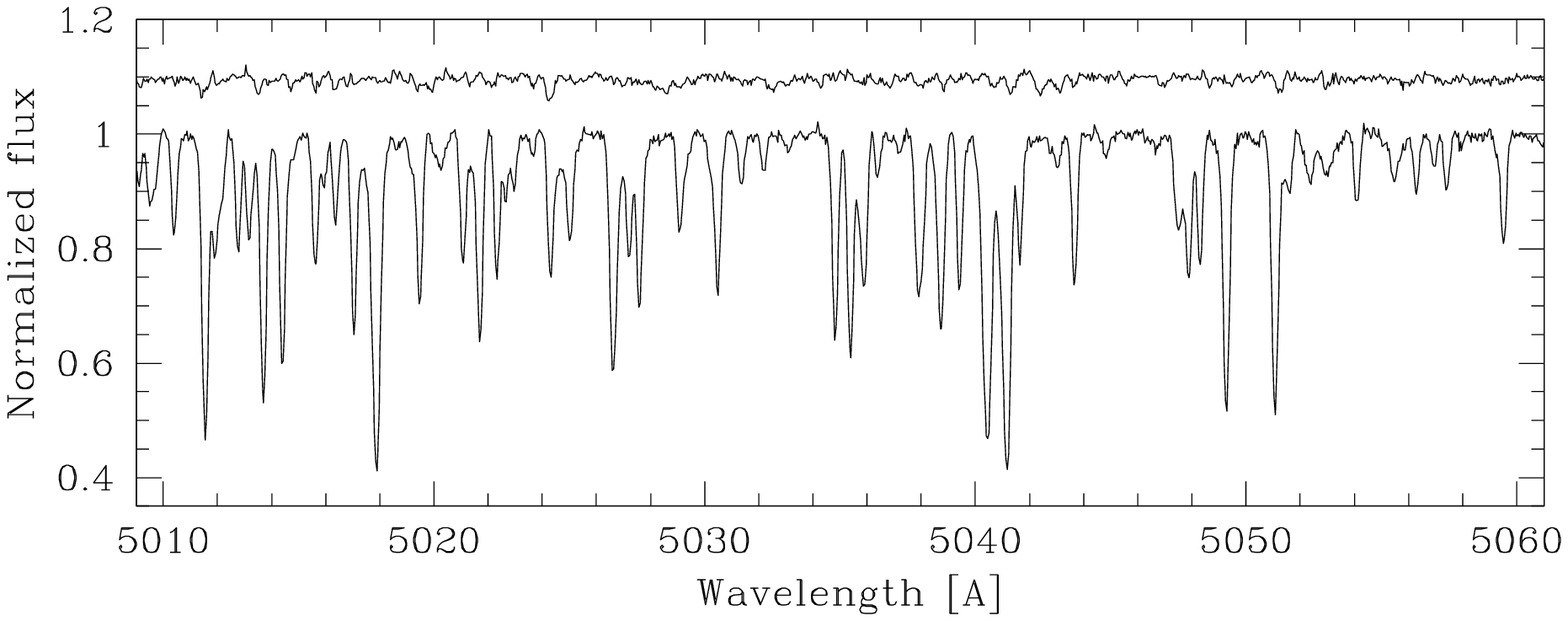}
\plotone{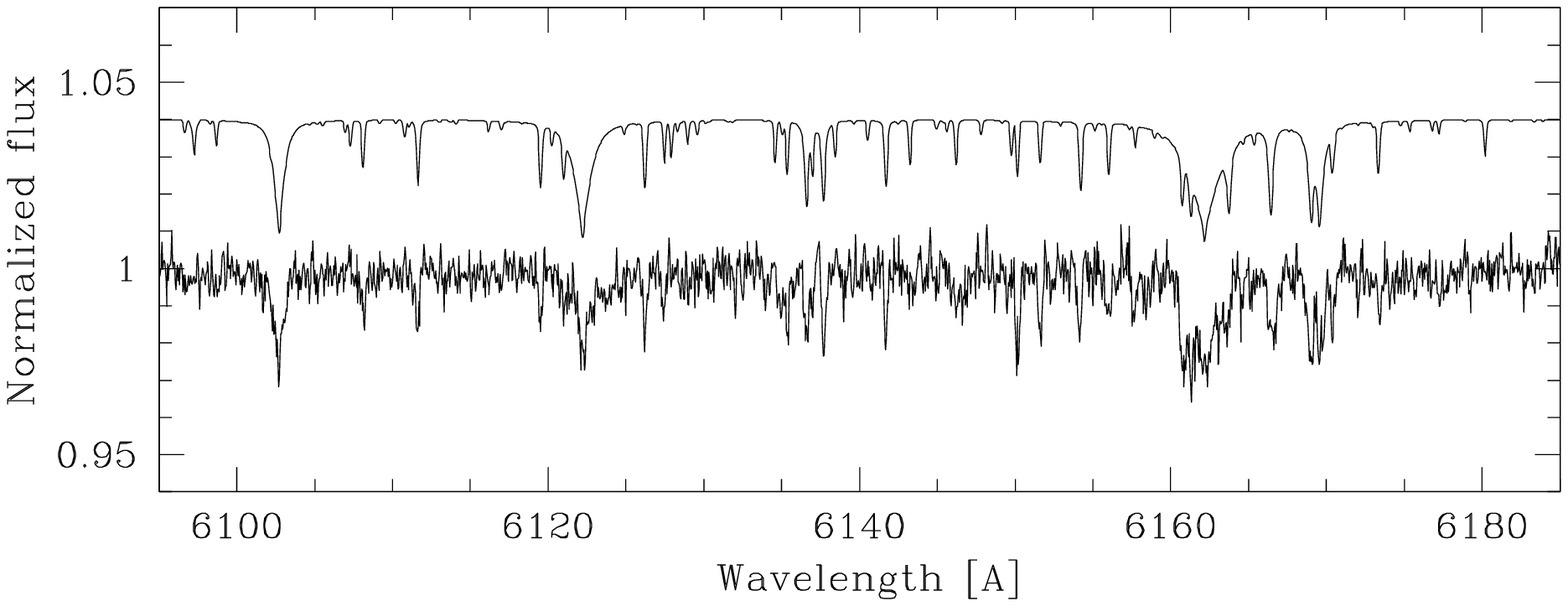}

\figcaption{\emph{Top:} Sample sections of the disentangled spectra of
  the primary and secondary of \vstar\ from our TRES
  observations. \emph{Bottom:} Disentangled spectrum of the secondary
  (bottom) in a region containing strong \ion{Ca}{1} lines, compared
  to a synthetic spectrum (top) with parameters $T_{\rm eff} =
  3900$\,K, $\log g = 4.65$, and $v \sin i = 5$\,\kms, close to those
  appropriate for the star. The model spectrum has been scaled to a
  light ratio of 4\% relative to the
  primary.\label{fig:disentangling}}

\end{figure}

The disentangled spectrum of the primary star gains in SNR compared to
the individual spectra roughly as ${\rm SNR_A} \sim \langle{\rm
  SNR}\rangle \sqrt{N}/ (1+\ell_{\rm B}/\ell_{\rm A})$, where $N$ is
the number of spectra and $\langle{\rm SNR}\rangle$ the average SNR of
the individual spectra. A similar expression holds for the
disentangled secondary spectrum, with the light ratio reversed. The
spectra resulting from the procedure have SNRs of 246 (primary) and 8
(secondary) for TRES ($\lambda$5800, $N = 27$), 103 and 1.4 for DS
($\lambda$5200, $N = 67$), and 713 and 30 for KPNO ($\lambda$6400, $N
= 30$). Portions of the disentangled TRES spectra appear in
Figure~\ref{fig:disentangling}, where a comparison with a model in the
bottom panel shows that the secondary spectrum was successfully
reconstructed from these observations, despite its faintness.

%%%%%%%%%%%%%%%%%%%%%%%%%%%%%%%%%%%%%%%%%%%%%%%%%%%%%%%%%%%%%%%%%%%%%%%%%%%
\section{Chemical abundance}
\label{sec:metallicity}
%%%%%%%%%%%%%%%%%%%%%%%%%%%%%%%%%%%%%%%%%%%%%%%%%%%%%%%%%%%%%%%%%%%%%%%%%%%

We subjected the disentangled spectra of the primary component to a
detailed analysis to determine the effective temperature and chemical
abundance. A first estimate of $T_{\rm eff}$ was made by fitting the
Balmer line profiles, which depend primarily on temperature and very
little on $\log g$, via genetic minimization \citep{Tamajo:11}.  Metal
lines in the wings were masked out, and the surface gravity and $v
\sin i$ were held fixed at values reported below in
Sect.~\ref{sec:dimensions}. We obtained temperatures of $5840 \pm
50$\,K and $5870 \pm 45$\,K from H$\alpha$ and H$\beta$ in the TRES
spectra, and $5780 \pm 55$\,K from H$\alpha$ in the KPNO
spectra. These uncertainties may be underestimated, however, as we
cannot rule out systematics from the normalization process and merging
of the echelle orders.

We then used the {\sc uclsyn} package \citep{Smalley:11} to fine-tune
the temperature and set the microturbulent velocity $\xi_{\rm t}$ from
the numerous \ion{Fe}{1} lines, and to determine the detailed
abundances based on the measured equivalent widths. Surface gravity
was held fixed as above. {\sc uclsyn} relies on synthetic spectra
computed under local thermodynamic equilibrium (LTE) using {\sc
  ATLAS9} model atmospheres \citep{Kurucz:79}. Excitation equilibrium
was imposed to determine $T_{\rm eff}$ from the \ion{Fe}{1} lines,
with the selection of lines and their $gf$ values taken from the
recent critical compilation of \cite{Bensby:14}. Microturbulence was
determined by enforcing no dependence between the abundances and the
reduced equivalent widths. We obtained $T_{\rm eff} = 5890 \pm 80$\,K
and $\xi_{\rm t} = 1.2 \pm 0.1$\,\kms\ from the TRES spectra, and
$T_{\rm eff} = 5970 \pm 110$\,K and $\xi_{\rm t} = 1.7 \pm
0.1$\,\kms\ from the red KPNO spectra. We attribute the discrepancy in
$\xi_{\rm t}$ values to the greatly different wavelength coverage of
the TRES and KPNO spectra.  The DS spectra do not permit independent
estimates of these parameters because of the very limited wavelength
coverage, so they were fixed at values of 5900\,K and 1.2\,\kms. We
collect the various temperature determinations for the primary star in
Table~\ref{tab:teff}, along with others described later, noting that
they are not all completely independent as some of them rely on the
same sets of spectra.

\begin{deluxetable}{lc}
\tablewidth{0pc}
\tablecaption{Effective temperature estimates for \vstar\,A.\label{tab:teff}}
\tablehead{
\colhead{~~~~~~~~~~~~~~~~~~~~Method~~~~~~~~~~~~~~~~~~~~} &
\colhead{$T_{\rm eff}$ (K)}
}
\startdata
TRES spectra, H$\alpha$\dotfill          &  $5840 \pm 50$\phn \\
TRES spectra, H$\beta$\dotfill           &  $5870 \pm 45$\phn \\
KPNO spectra, H$\alpha$\dotfill          &  $5780 \pm 55$\phn \\
TRES spectra, {\sc uclsyn}\dotfill       &  $5890 \pm 80$\phn \\
KPNO spectra, {\sc uclsyn}\dotfill       &  $5970 \pm 110$ \\
DS spectra, cross-correlation\dotfill    &  $5880 \pm 100$ \\
TRES spectra, cross-correlation\dotfill  &  $5880 \pm 100$ \\
KPNO spectra, cross-correlation\dotfill  &  $5820 \pm 100$ \\
Color indices and $J_{\rm B}$\dotfill    &  $5950 \pm 30$\phn
\enddata

\tablecomments{Uncertainties are formal errors, and may not
  reflect systematics.}

\end{deluxetable}

Detailed abundances on the scale of \cite{Asplund:09} were obtained
for 21 species from the TRES spectra, as listed in
Table~\ref{tab:met_TRES.DS.KPNO}, and somewhat fewer for the DS and
KPNO spectra. The uncertainties account for errors in $T_{\rm eff}$
and $\xi_{\rm t}$ of 100\,K and 0.1\,\kms, respectively.  The
agreement between the three instruments is excellent, the average
differences for all elements taken together being $\langle{\rm
  TRES}-{\rm DS}\rangle = +0.022 \pm 0.014$ dex (10 lines in common),
$\langle{\rm TRES}-{\rm KPNO}\rangle = -0.011 \pm 0.032$ dex (7
lines), and $\langle{\rm DS}-{\rm KPNO}\rangle = -0.022 \pm 0.029$ dex
(4 lines). In particular, the iron abundances based on \ion{Fe}{1} are
very consistent. Those from \ion{Fe}{2} are somewhat less reliable and
are based on far fewer lines. We adopted the weighted average of the
\ion{Fe}{1} values, ${\rm [Fe/H]} = -0.12 \pm 0.08$, with a
conservative uncertainty.  Abundances of most other elements in
\vstar\ tend to be subsolar as well. This includes the $\alpha$
elements, which are therefore not enhanced in this system.

\begin{deluxetable*}{ll c cc c cc c cc c c}
\tablewidth{0pc}
\tablecaption{\vstar\ abundances from our disentangled TRES, DS, and KPNO spectra.\label{tab:met_TRES.DS.KPNO}}
\tablehead{
\colhead{} &
\colhead{} & &
\multicolumn{2}{c}{TRES} & &
\multicolumn{2}{c}{DS}   & &
\multicolumn{2}{c}{KPNO} & &
\colhead{} \\
\cline{4-5} \cline{7-8} \cline{10-11} \\ [-1ex]
\colhead{A} &
\colhead{Elem} & &
\colhead{$N$} &
\colhead{[X/H]} & &
\colhead{$N$} &
\colhead{[X/H]} & &
\colhead{$N$} &
\colhead{[X/H]} & &
\colhead{$\log\epsilon_{\sun}$}
}
\startdata
 6 &  \ion{C}{1}  &&    4  &  $+0.06 \pm 0.10$  &&\nodata&    \nodata        &&\nodata&    \nodata       &&   $8.43 \pm 0.05$ \\
11 &  \ion{Na}{1} &&    5  &  $-0.07 \pm 0.09$  &&\nodata&    \nodata        &&\nodata&    \nodata       &&   $6.24 \pm 0.04$ \\
12 &  \ion{Mg}{1} &&    9  &  $-0.16 \pm 0.07$  &&   3   & $-0.24 \pm 0.09$  &&  3    & $-0.29 \pm 0.06$ &&   $7.60 \pm 0.04$ \\
13 &  \ion{Al}{1} &&    4  &  $-0.06 \pm 0.08$  &&\nodata&    \nodata        &&\nodata&    \nodata       &&   $6.45 \pm 0.03$ \\
14 &  \ion{Si}{1} &&   15  &  $-0.11 \pm 0.04$  &&\nodata&    \nodata        && 10    & $-0.04 \pm 0.06$ &&   $7.51 \pm 0.03$ \\
16 &  \ion{S}{1}  &&    5  &  $+0.02 \pm 0.10$  &&\nodata&    \nodata        &&\nodata&    \nodata       &&   $7.12 \pm 0.03$ \\
20 &  \ion{Ca}{1} &&   21  &  $-0.03 \pm 0.10$  &&\nodata&    \nodata        &&  9    & $-0.11 \pm 0.09$ &&   $6.34 \pm 0.04$ \\
21 &  \ion{Sc}{2} &&   12  &  $-0.11 \pm 0.06$  &&\nodata&    \nodata        &&\nodata&    \nodata       &&   $3.15 \pm 0.04$ \\
22 &  \ion{Ti}{1} &&   32  &  $-0.10 \pm 0.10$  &&   5   & $-0.12 \pm 0.12$  &&\nodata&    \nodata       &&   $4.95 \pm 0.05$ \\
23 &  \ion{V}{1}  &&   28  &  $+0.04 \pm 0.11$  &&   4   & $+0.07 \pm 0.12$  &&\nodata&    \nodata       &&   $3.93 \pm 0.08$ \\
24 &  \ion{Cr}{1} &&   15  &  $-0.12 \pm 0.09$  &&  19   & $-0.07 \pm 0.07$  &&  2    & $-0.05 \pm 0.08$ &&   $5.64 \pm 0.04$ \\
25 &  \ion{Mn}{1} &&   19  &  $-0.09 \pm 0.09$  &&   2   & $-0.14 \pm 0.13$  &&\nodata&    \nodata       &&   $5.43 \pm 0.05$ \\
26 &  \ion{Fe}{1} &&  132  &  $-0.11 \pm 0.06$  &&  38   & $-0.14 \pm 0.09$  && 41    & $-0.11 \pm 0.07$ &&   $7.50 \pm 0.04$ \\
26 &  \ion{Fe}{2} &&   23  &  $-0.16 \pm 0.08$  &&\nodata&    \nodata        &&  4    & $-0.07 \pm 0.06$ &&   $7.50 \pm 0.04$ \\
27 &  \ion{Co}{1} &&   11  &  $-0.12 \pm 0.09$  &&   7   & $-0.18 \pm 0.09$  &&\nodata&    \nodata       &&   $4.99 \pm 0.07$ \\
28 &  \ion{Ni}{1} &&   48  &  $-0.13 \pm 0.09$  &&  13   & $-0.16 \pm 0.10$  && 12    & $-0.07 \pm 0.06$ &&   $6.22 \pm 0.04$ \\
29 &  \ion{Cu}{1} &&    4  &  $-0.15 \pm 0.06$  &&\nodata&    \nodata        &&\nodata&    \nodata       &&   $4.19 \pm 0.04$ \\
30 &  \ion{Zn}{1} &&    3  &  $-0.23 \pm 0.09$  &&\nodata&    \nodata        &&\nodata&    \nodata       &&   $4.56 \pm 0.05$ \\
39 &  \ion{Y}{2}  &&   10  &  $-0.28 \pm 0.07$  &&   3   & $-0.34 \pm 0.10$  &&\nodata&    \nodata       &&   $2.21 \pm 0.05$ \\
56 &  \ion{Ba}{2} &&    5  &  $-0.18 \pm 0.12$  &&\nodata&    \nodata        &&\nodata&    \nodata       &&   $2.18 \pm 0.09$ \\
60 &  \ion{Nd}{2} &&   10  &  $-0.03 \pm 0.06$  &&   4   & $+0.00 \pm 0.07$  &&\nodata&    \nodata       &&   $1.42 \pm 0.04$
\enddata

\tablecomments{Columns list the atomic number, the element and
  ionization degree, the number of spectral lines measured and
  abundance relative to the Sun from each instrument, and finally the
  reference photospheric solar values from \cite{Asplund:09}.
  Abundances of other elements based on a single line are considered
  less reliable and are not listed.}

\end{deluxetable*}

%%%%%%%%%%%%%%%%%%%%%%%%%%%%%%%%%%%%%%%%%%%%%%%%%%%%%%%%%%%%%%%%%%%%%%%%%%%
\section{Photometric observations}
\label{sec:photometry}
%%%%%%%%%%%%%%%%%%%%%%%%%%%%%%%%%%%%%%%%%%%%%%%%%%%%%%%%%%%%%%%%%%%%%%%%%%%

Two sets of $V$-band images of \vstar\ were obtained with independent
robotic telescopes operating at the University of Arkansas (URSA
WebScope) and near Silver City, NM (NFO WebScope) from 2001 January to
2012 February. A description of the telescopes and instrumentation, as
well as the data acquisition and reduction procedures may be found in
the papers by \cite{Grauer:08} and \cite{Lacy:12}. We collected a
total of 5137 URSA observations and 3024 NFO observations providing
complete phase coverage. The comparison (`comp') and check (`ck')
stars were HD\,294597 (TYC\,4786-1469-1; $V = 10.43$) and HD\,294593
(TYC\,4786-2281-1; $V = 9.56$). The differential URSA measurements (in
the sense variable minus comp) are listed in Table~\ref{tab:ursa};
those from the NFO appear in Table~\ref{tab:nfo} (computed as variable
minus `comps', where comps is the magnitude corresponding to the sum
of the fluxes of the comp and ck stars). The precision of these
measurements is about 7 milli-magnitudes (mmag) for URSA and 5 mmag
for NFO. A graphical representation of these observations is shown
later in Sect.~\ref{sec:lightcurves}.

\begin{deluxetable}{lcc}
\tablewidth{0pc}
\tablecaption{Differential $V$-band measurements of \vstar\ from URSA.\label{tab:ursa}}
\tablehead{
\colhead{HJD} &
\colhead{} &
\colhead{$\Delta V$} \\
\colhead{($2,\!400,\!000+$)} &
\colhead{Phase\tablenotemark{a}} &
\colhead{(mag)}
}
\startdata
  51929.75550  &  0.5421  &  0.676 \\
  51929.75652  &  0.5423  &  0.681 \\
  51929.75754  &  0.5425  &  0.676 \\
  51929.75855  &  0.5426  &  0.678 \\
  51929.75957  &  0.5428  &  0.679
\enddata

\tablenotetext{a}{Phase counted from the reference epoch of primary
  eclipse given in Sect.~\ref{sec:ephemeris}.}

\tablecomments{This table is available in its entirety in
  machine-readable and Virtual Observatory (VO) forms in the online
  journal. A portion is shown here for guidance regarding its form and
  content.}
\end{deluxetable}

\begin{deluxetable}{lcc}
\tablewidth{0pc}
\tablecaption{Differential $V$-band measurements of \vstar\ from NFO.\label{tab:nfo}}
\tablehead{
\colhead{HJD} &
\colhead{} &
\colhead{$\Delta V$} \\
\colhead{($2,\!400,\!000+$)} &
\colhead{Phase\tablenotemark{a}} &
\colhead{(mag)}
}
\startdata
  53377.63997  &  0.4816  &  0.671 \\
  53377.64133  &  0.4818  &  0.674 \\
  53377.64273  &  0.4820  &  0.674 \\
  53377.64415  &  0.4823  &  0.674 \\
  53377.64551  &  0.4825  &  0.673
\enddata

\tablenotetext{a}{Phase counted from the reference epoch of primary
  eclipse given in Sect.~\ref{sec:ephemeris}.}

\tablecomments{This table is available in its entirety in
  machine-readable and Virtual Observatory (VO) forms in the online
  journal. A portion is shown here for guidance regarding its form and
  content.}
\end{deluxetable}

\begin{figure}
\epsscale{1.15}
\plotone{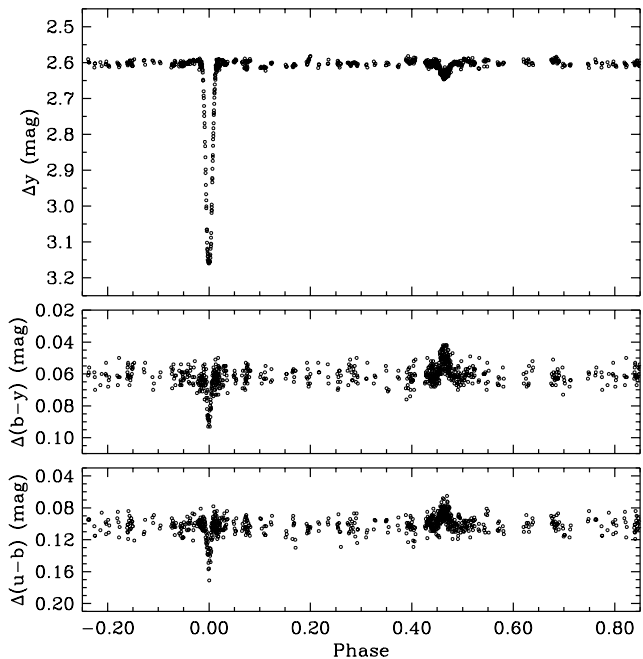}

\figcaption{Differential Str\"omgren photometry of \vstar.\label{fig:uvby}}

\end{figure}

Differential photometric measurements of \vstar\ were also gathered
with the Str\"omgren Automatic Telescope at ESO (La Silla, Chile),
during several campaigns from 2001 January to 2006 February. A total
of 720 observations were made in the $uvby$ bands, using the three
comparison stars HD\,39438 (\ion{F5}{5}), HD\,39833 (\ion{G0}{3}), and
HD\,40590 (\ion{F6}{5}). The typical precision per differential
measurement ranges from 7 mmag in $y$ to 11 mmag in $u$, and the phase
coverage is complete. The reduction of this material followed
procedures analogous to those described by \cite{Clausen:08}. We
report these observations in Table~\ref{tab:uvby}, and show them
graphically in Figure~\ref{fig:uvby}. In addition to the light curves,
we obtained homogeneous standard $uvby\beta$ indices with the same
telescope on dedicated nights in which \vstar\ and the comparison
stars were observed together with a large sample of standard
stars. The resulting indices outside of eclipse are $V = 9.861 \pm
0.008$, $b-y = 0.408 \pm 0.005$, $m_1 = 0.199 \pm 0.009$, $c_1 = 0.296
\pm 0.010$, and $\beta = 2.589 \pm 0.007$.

\begin{deluxetable}{lccccc}
\tablewidth{0pc}
\tablecaption{Differential $uvby$ measurements of \vstar.\label{tab:uvby}}
\tablehead{
\colhead{HJD} &
\colhead{} &
\colhead{$\Delta u$} &
\colhead{$\Delta v$} &
\colhead{$\Delta b$} &
\colhead{$\Delta y$} \\
\colhead{($2,\!400,\!000+$)} &
\colhead{Phase\tablenotemark{a}} &
\colhead{(mag)} &
\colhead{(mag)} &
\colhead{(mag)} &
\colhead{(mag)}
}
\startdata
 51929.59746  & 0.5162  &  2.766  &  2.758  &  2.668  &  2.608 \\
 51929.60318  & 0.5172  &  2.759  &  2.757  &  2.666  &  2.604 \\
 51929.60786  & 0.5179  &  2.766  &  2.756  &  2.662  &  2.602 \\
 51929.61784  & 0.5196  &  2.755  &  2.749  &  2.653  &  2.587 \\
 51929.62253  & 0.5203  &  2.763  &  2.750  &  2.655  &  2.600
\enddata

\tablenotetext{a}{Phase counted from the reference epoch of primary
  eclipse given in Sect.~\ref{sec:ephemeris}.}

\tablecomments{This table is available in its entirety in
  machine-readable and Virtual Observatory (VO) forms in the online
  journal. A portion is shown here for guidance regarding its form and
  content.}
\end{deluxetable}

\begin{figure}
\epsscale{1.15}
\plotone{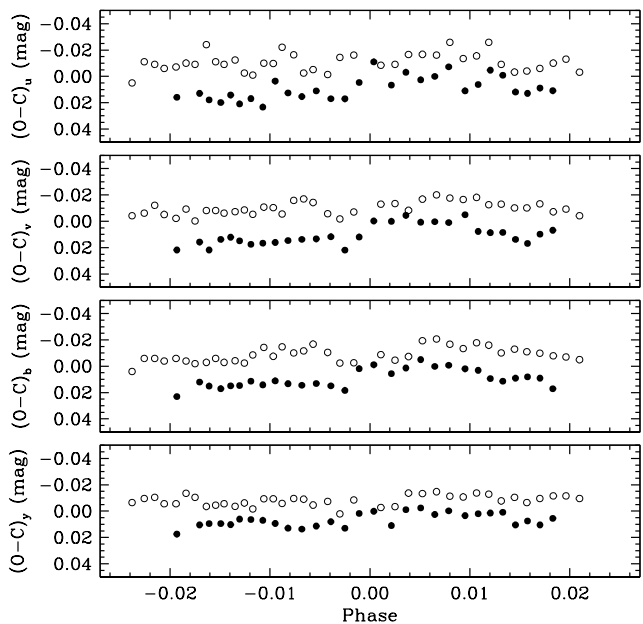}

\figcaption{Residuals of the differential Str\"omgren photometry of
  \vstar\ from the light curve fits described in
  Sect.~\ref{sec:lightcurves}, shown for two separate nights: JD
  $2,\!452,\!989$ (open symbols) and JD $2,\!452,\!604$ (filled
  symbols). The offset of $\sim$0.02 mag is likely due to spottedness
  on the primary.\label{fig:spots}}

\end{figure}

Close examination of the photometry shows clear night-to-night
variations that appear to be intrinsic to the system and are likely
due to star spots, presumably on the much brighter primary. This would
be consistent with the signs of activity noted previously. An illustration
of this is seen in Figure~\ref{fig:spots}, in which instead of the
original data we show for clarity the residuals of the $uvby$
measurements near the primary eclipse from the photometric solutions
described in the next section. Two different nights are represented
with different symbols (open circles for JD $2,\!452,\!989$, filled
circles for $2,\!452,\!604$), and display an offset of $\sim$0.02
mag. Similar offsets are seen at other orbital phases.

%%%%%%%%%%%%%%%%%%%%%%%%%%%%%%%%%%%%%%%%%%%%%%%%%%%%%%%%%%%%%%%%%%%%%%%%%%%
\section{Light curve analysis}
\label{sec:lightcurves}
%%%%%%%%%%%%%%%%%%%%%%%%%%%%%%%%%%%%%%%%%%%%%%%%%%%%%%%%%%%%%%%%%%%%%%%%%%%

The $V$-band and $uvby$ data of \vstar\ were analyzed using the
JKTEBOP code of John Southworth \citep{Nelson:72, Popper:81,
  Southworth:04}, which is adequate for relatively uncomplicated
systems such as this that are well detached. The fitted light-curve
parameters are the central surface brightness of the smaller, fainter,
cooler, and less massive star (secondary) relative to the other
($J_{\rm B}$), the sum of the relative radii of the primary and
secondary in units of the semi-major axis ($r_{\rm A} + r_{\rm B}$),
the radius ratio ($k \equiv r_{\rm B}/r_{\rm A}$), the inclination
angle of the orbit ($i$), the orbital eccentricity and longitude of
periastron of the primary ($e$ and $\omega$), and the linear
limb-darkening coefficients ($u_{\rm A}$ and $u_{\rm B}$). The
ephemeris used in the solutions was that of Sect.~\ref{sec:ephemeris},
and the mass ratio was held fixed at the spectroscopic value $q =
0.5932$.  Because the secondary eclipse is so shallow, the
limb-darkening parameters for the smaller star were fixed at
theoretical values based on an average of predictions from
\cite{VanHamme:93}, \cite{Diaz-Cordoves:95}, \cite{Claret:00}, and
\cite{Claret:03}, and the values for the larger star were allowed to
vary. Gravity darkening exponents based on the components'
temperatures were taken from theory \citep{Claret:98}. The light curve
modeling was carried out using the Levenberg-Marquardt option in
JKTEBOP, but the results and their uncertainties were checked by
performing a Monte Carlo simulation study, and found to agree well
between the two methods.

\begin{figure}
\epsscale{1.15} 
\plotone{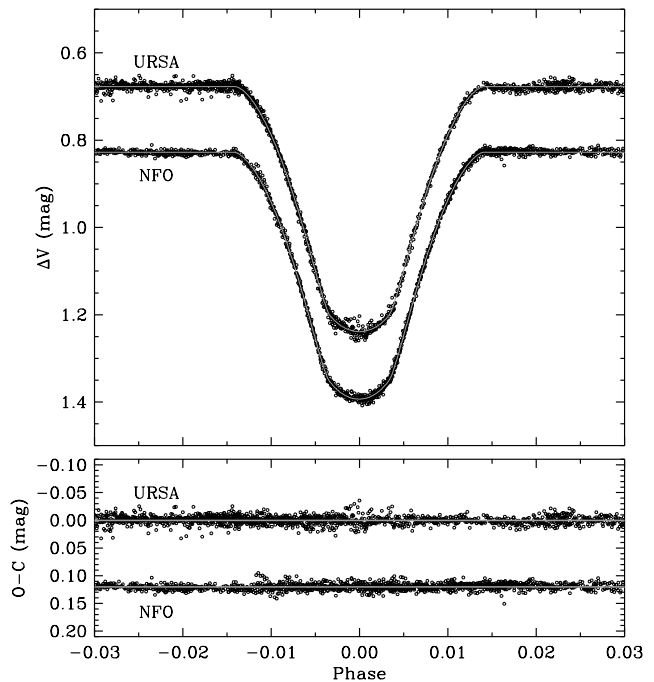} 
\figcaption{URSA and NFO differential $V$-band observations of
  \vstar\ near primary eclipse, shown with our best model
  fits. Residuals from the fits are shown at the bottom, with those
  from NFO displaced vertically for clarity.\label{fig:Vprim}}
\end{figure}

\begin{figure}
\epsscale{1.15} 
\plotone{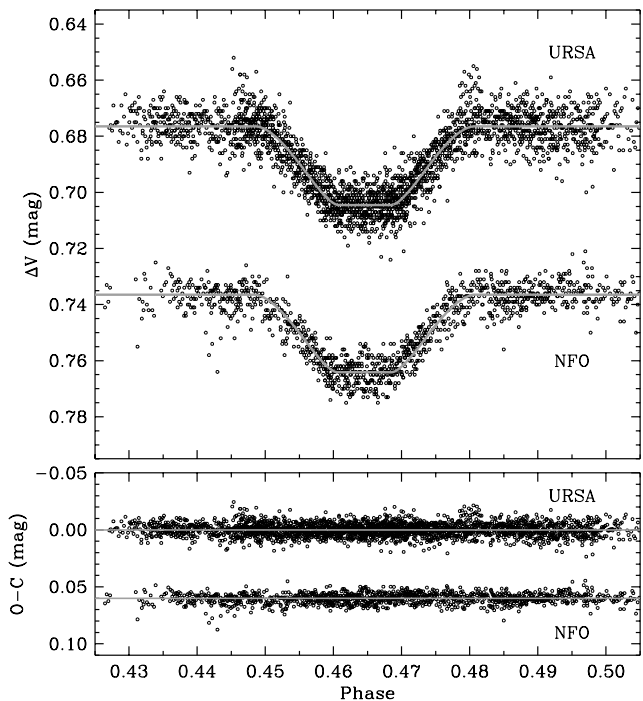} 

\figcaption{URSA and NFO differential $V$-band observations of
  \vstar\ near secondary eclipse, shown with our best model fits. Note
  the different vertical scale compared to Figure~\ref{fig:Vprim}.
  Residuals from the fits are shown at the bottom, with those from NFO
  displaced vertically for clarity. The large open circles on the
  ascending branch mark the time of the ROSAT X-ray observation
  described in Sect.~\ref{sec:activity}.\label{fig:Vsec}}

\end{figure}

\begin{figure}
\epsscale{1.10} 
\plotone{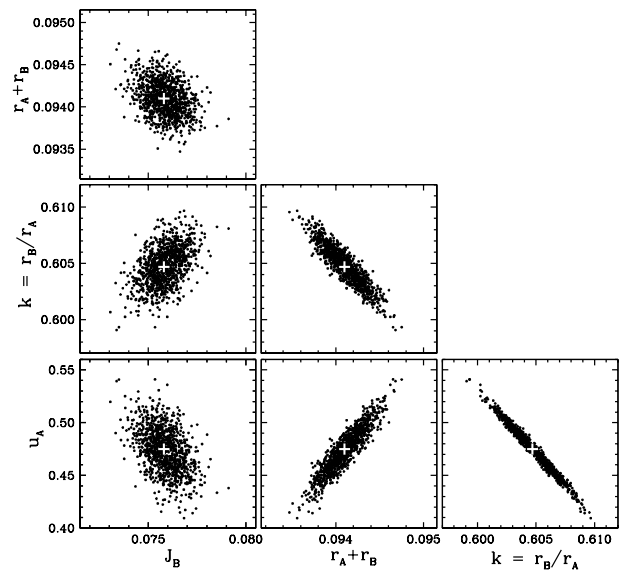} 

\figcaption{Results from Monte Carlo simulations with JKTEBOP using
  the URSA data set, illustrating the correlations among some of the
  main elements: $J_{\rm B}$, $r_{\rm A}+r_{\rm B}$, $k \equiv r_{\rm
    B}/r_{\rm A}$, and the primary linear limb-darkening parameter
  $u_{\rm A}$. Plus signs represent the median values for each
  variable.\label{fig:correlation}}

\end{figure}

Preliminary fits showed that the values for $i$, $e$, and $\omega$
were very consistent among the data sets, so weighted mean values were
adopted ($i = 89\fdg78 \pm 0\fdg08$, $e = 0.0862 \pm 0.0010$, $\omega
= 130\fdg08 \pm 0\fdg14$) and held fixed for the final solutions. The
results for the different data sets are presented in
Table~\ref{tab:LCfits}, where $\ell_{\rm A}$ and $\ell_{\rm B}$ are
the light fractions of the components at orbital quadrature, $\sigma$
is the rms residual in mmag, and $N$ is the number of observations.
The fits for the URSA and NFO data near the primary and secondary
eclipses are illustrated in Figure~\ref{fig:Vprim} and
Figure~\ref{fig:Vsec}, respectively. An illustration of the
correlation between some of the main variables is shown in
Figure~\ref{fig:correlation}, based on a Monte Carlo simulation with
1000 trials using the URSA data set.

\begin{deluxetable*}{lccccccc}
\tablewidth{0pt}
\tablecaption{Light curve solutions for \vstar.\label{tab:LCfits}}
\tablehead{
\colhead{~~~~Parameter~~~~} &
\colhead{$u$} & 
\colhead{$v$} & 
\colhead{$b$} & 
\colhead{$y$} & 
\colhead{URSA $V$} & 
\colhead{NFO $V$} &
\colhead{Adopted}
}
\startdata
$J_{\rm B}$\dotfill                    &  0.0066  & 0.0200  & 0.0537  & 0.0867 &  0.0758  &  0.0739  &   \phm{\tablenotemark{a}}$0.075 \pm 0.002$\tablenotemark{a} \\
$r_{\rm A}+r_{\rm B}$\dotfill          &  0.0971  & 0.0960  & 0.0964  & 0.0956 &  0.0941  &  0.0950  &   $0.0953 \pm 0.0010$  \\
$r_{\rm A}$\dotfill                    &  0.0615  & 0.0602  & 0.0604  & 0.0598 &  0.0587  &  0.0594  &   $0.0596 \pm 0.0008$  \\
$r_{\rm B}$\dotfill                    &  0.0356  & 0.0358  & 0.0360  & 0.0358 &  0.0354  &  0.0357  &   $0.0357 \pm 0.0004$  \\
$k \equiv r_{\rm B}/r_{\rm A}$\dotfill &  0.578   & 0.595   & 0.596   & 0.599  &  0.604   &  0.600   &   $0.600 \pm 0.004$    \\
$i$ (deg)\dotfill                      &  89.82\tablenotemark{b}  & 89.63   & 89.67   & 89.82\tablenotemark{b} &  89.82\tablenotemark{b}  &  89.80   &  $89.78 \pm 0.08$\phn     \\
$e$\dotfill                            &  0.0779  & 0.0801  & 0.0851  & 0.0862 &  0.0863  &  0.0870  &   $0.0862 \pm 0.0010$  \\
$\omega_{\rm A}$ (deg)\dotfill         &  130.26  & 130.11  & 130.19  & 129.94 &  130.10  &  130.08  &  $130.08 \pm 0.14$\phn\phn  \\
$u_{\rm A}$\dotfill                    &  0.92    & 0.75    & 0.64    & 0.52   &  0.48    &  0.54    &     \nodata  \\
$u_{\rm B}$\dotfill                    &  0.78\tablenotemark{b}   & 0.75\tablenotemark{b}   & 0.79\tablenotemark{b}   & 0.72\tablenotemark{b}  &  0.71\tablenotemark{b}   &  0.71\tablenotemark{b}   &     \nodata  \\
$\ell_{\rm A}$\dotfill                 &  0.9974  & 0.9928  & 0.9822  & 0.9720 &  0.9753  &  0.9756  &     \nodata  \\
$\ell_{\rm B}$\dotfill                 &  0.0024  & 0.0070  & 0.0176  & 0.0278 &  0.0245  &  0.0242  &     \nodata  \\
$\ell_{\rm B}/\ell_{\rm A}$\dotfill    &  0.002   & 0.007   & 0.018   & 0.029  &  0.025   &  0.025   &     \nodata  \\
$\sigma$ (mmag)\dotfill                &  10.960  & 8.700   & 7.821   & 7.165  &  6.678   &  5.131   &     \nodata  \\
$N$\dotfill                            &   720    & 720     & 720     & 720    &  5137    &  3024    &     \nodata  
\enddata
\tablenotetext{a}{Average value for the $V$ band, with a conservative uncertainty.}
\tablenotetext{b}{Held fixed.}
\end{deluxetable*}

Our solutions consistently indicate that the secondary eclipse (only
0.028 mag deep in $V$) is total, with a duration of totality of about
70 minutes. The primary eclipse is annular.  Trials were made allowing
for the possible presence of third light, but the resulting values
were not significantly different from zero, so no third light was
allowed in the final solutions.  Additional trials were carried out
using a non-linear limb-darkening law of the logarithmic type
\citep{Claret:00}, and also a quadratic law, but we found the residual
variances of the fits to be always worse than with the linear
limb-darkening law.  The resulting fitted orbital parameters were not
significantly different from those with the linear law, except that
the logarithmic law preferred a primary relative radius value ($r_{\rm
  A}$) about 1\% larger, and the quadratic law gave a value about
1.9\% larger.  Because the fit to the data is superior for the linear
law, we have chosen those results for the remainder of this
study. Average values of the geometric properties used for computing
the absolute dimensions are listed in the last column of
Table~\ref{tab:LCfits}.

%%%%%%%%%%%%%%%%%%%%%%%%%%%%%%%%%%%%%%%%%%%%%%%%%%%%%%%%%%%%%%%%%%%%%%%%%%%
\section{Absolute dimensions}
\label{sec:dimensions}
%%%%%%%%%%%%%%%%%%%%%%%%%%%%%%%%%%%%%%%%%%%%%%%%%%%%%%%%%%%%%%%%%%%%%%%%%%%

Masses and radii for the components of \vstar\ computed from the
information in Table~\ref{tab:orbits} and Table~\ref{tab:LCfits} are
presented in Table~\ref{tab:absdim}, and are determined to better than
0.7\% in the case of the masses and 1.3\% for the radii. Based on the
three detailed and independent chemical analyses in
Sect.~\ref{sec:metallicity}, the average metallicity of
\vstar\ (assuming the primary and secondary to have the same
composition) is determined to be ${\rm [Fe/H]} = -0.12 \pm 0.08$. A
photometric estimate in good agreement with this value was obtained
using the Str\"omgren indices in Sect.~\ref{sec:photometry}
weight-averaged with those measured by \cite{Lacy:02}, along with the
calibration in Eq.~14 by \cite{Olsen:84}. The result is ${\rm [Fe/H]}
= -0.10 \pm 0.13$, which should be unaffected by the very faint
secondary. Use of the calibration by \cite{Holmberg:07} yields a
somewhat lower value of ${\rm [Fe/H]} = -0.23 \pm 0.09$, still in
agreement with the more reliable spectroscopic determination.

The procedure described in Sect.~\ref{sec:spectroscopy} to determine
template parameters for deriving RVs can be refined by interpolating
between grid points in our libraries of synthetic spectra, in order to
determine more precise values for $T_{\rm eff}$ and $v \sin i$. The $v
\sin i$ value for the primary obtained in this way, $9 \pm 1$\,\kms,
is consistent with what is expected if the star were rotating
pseudo-synchronously \citep[see Table~\ref{tab:absdim};][]{Hut:81},
and is in agreement with predictions from theory suggesting a
synchronization timescale of only $\sim$10$^7$\,yr
(Sect.~\ref{sec:spectroscopy}), much shorter than the system age
estimated below.  However, the resulting temperature for that star
from this method depends on the metallicity adopted, due to strong
correlations between those two properties. We performed the
determinations with [Fe/H] values of 0.0 and $-0.5$, and then
interpolated to ${\rm [Fe/H]} = -0.12$, separately for our DS, TRES,
and KPNO spectra. The $T_{\rm eff}$ values obtained for the primary
are 5880\,K, 5880\,K, and 5820\,K, respectively, which are similar to
those derived from disentangling (Sect.~\ref{sec:metallicity}). They
have estimated uncertainties of 100\,K. The accuracy of our various
(non-independent) temperature determinations for the primary star,
which we have summarized in Table~\ref{tab:teff}, is likely limited by
systematic effects not reflected in the formal uncertainties. For the
analysis that follows we have adopted a consensus temperature for the
primary of $5890 \pm 100$\,K, in which the uncertainty is a
conservative estimate that is approximately equal to half the spread
in the spectroscopic determinations. The secondary temperature was
inferred from this value and the temperature difference, $\Delta
T_{\rm eff}$. The latter may be derived from the central surface
brightness ratio $J_{\rm B}$ (Table~\ref{tab:LCfits}) using the
absolute visual flux calibration of \cite{Popper:80}. As this
procedure is entirely differential, the resulting temperature
difference, $\Delta T_{\rm eff} = 2010 \pm 70$\,K is typically better
determined than the individual temperatures. The adopted $T_{\rm eff}$
value for the secondary is then $3880 \pm 120$\,K. These stellar
temperatures correspond approximately to spectral types of G1 and M1
for the primary and secondary. We note, finally, that the small
differences between these final stellar properties and the template
parameters adopted in Sect.~\ref{sec:spectroscopy} for the RV
determinations have a negligible effect on those measurements.

\begin{deluxetable}{lcc}
\tablewidth{0pt}
\tablecaption{Physical properties of \vstar.\label{tab:absdim}}
\tablehead{
\colhead{~~~~~~~~~Parameter~~~~~~~~~} &
\colhead{Star A} & 
\colhead{Star B}
}
\startdata
Mass ($M_{\sun}$)\dotfill     & 1.0038~$\pm$~0.0066      & 0.5955~$\pm$~0.0022   \\
Radius ($R_{\sun}$)\dotfill   &  0.980~$\pm$~0.013       & 0.5873~$\pm$~0.0067   \\
$\log g$ (cgs)\dotfill        &  4.457~$\pm$~0.012       &  4.676~$\pm$~0.010    \\
$T_{\rm eff}$ (K)\dotfill     &   5890~$\pm$~100\phn     &   3880~$\pm$~120\phn  \\
$\Delta T_{\rm eff}$ (K)\dotfill      &      \multicolumn{2}{c}{$2010~\pm~70$\phn\phn}  \\
$a$ ($R_{\sun}$)\dotfill      &      \multicolumn{2}{c}{16.450~$\pm$~0.030\phn}       \\ [-0.5ex]
$v_{\rm sync} \sin i$ (\kms)\tablenotemark{a}\dotfill &    8.1~$\pm$~0.1    &    4.9~$\pm$~0.1  \\
$v_{\rm psync} \sin i$ (\kms)\tablenotemark{b}\dotfill &    8.5~$\pm$~0.1    &    5.1~$\pm$~0.1  \\
$v \sin i$ (\kms)\tablenotemark{c}\dotfill    &      9~$\pm$~1  &     \nodata           \\
$\log L/L_\sun$\dotfill       &  0.016~$\pm$~0.032       & $-$1.154~$\pm$~0.053\phs    \\
$L_{\rm B}/L_{\rm A}$\dotfill         &      \multicolumn{2}{c}{0.068~$\pm$~0.009}        \\
$M_{\rm bol}$ (mag)\dotfill   &  4.693~$\pm$~0.079       &  7.62~$\pm$~0.13    \\
$F_V$\tablenotemark{d}\dotfill       & 3.7586~$\pm$~0.0098      & 3.468~$\pm$~0.011    \\
$M_V$ (mag)\tablenotemark{d}\dotfill &   4.71~$\pm$~0.10        &   8.72~$\pm$~0.11     \\
$E(B-V)$ (mag)\dotfill        &      \multicolumn{2}{c}{0.045~$\pm$~0.020}       \\
$V-M_V$ (mag)\tablenotemark{d}\dotfill        &      \multicolumn{2}{c}{5.06~$\pm$~0.12}        \\
Distance (pc)\tablenotemark{d}\dotfill        &      \multicolumn{2}{c}{103~$\pm$~6\phn\phn}    \\
${\rm [Fe/H]}$\dotfill                        & \multicolumn{2}{c}{$-0.12 \pm 0.08$\phs}
\enddata
\tablenotetext{a}{Projected rotational velocity assuming synchronous rotation with the mean
orbital motion.}
\tablenotetext{b}{Projected rotational velocity assuming pseudo-synchronous rotation.}
\tablenotetext{c}{Value measured spectroscopically.}
\tablenotetext{d}{Relies on the absolute visual flux ($F_V$) calibration of \cite{Popper:80}.}
\end{deluxetable}

The reddening towards \vstar\ was estimated in several ways. One comes
from the Str\"omgren photometry and the calibration by
\cite{Crawford:75}, and gives $E(B-V) = 0.059$. Five other $E(B-V)$
values were inferred from the extinction maps of \cite{Burstein:82},
\cite{Hakkila:97}, \cite{Schlegel:98}, \cite{Drimmel:03}, and
\cite{Amores:05} for an assumed distance of 100\,pc. The results,
0.071, 0.039, 0.052, 0.019, and 0.030, were averaged with the previous
one to yield an adopted reddening of $E(B-V) = 0.045 \pm 0.020$, with
a conservative uncertainty. A consistency check on the effective
temperature adopted above may be obtained from standard photometry
available for \vstar\ from various catalogs and other literature
sources (Tycho-2, \citealt{Hog:00}; 2MASS, \citealt{Cutri:03}; TASS,
\citealt{Droege:06}; APASS, \citealt{Henden:12}; \citealt{Lacy:92b};
\citealt{Lacy:02}; and Sect.~\ref{sec:photometry}). From eleven
appropriately de-reddened non-independent color indices and the
calibrations of \cite{Casagrande:10} (for the above adopted
spectroscopic metallicity) we obtained $T_{\rm eff} = 5800 \pm
100$\,K, which corresponds to the combined light of the two stars as
the secondary has a non-negligible influence on the photometry,
especially at the redder wavelengths.  Individual temperatures for the
components may then be inferred using the absolute visual flux
calibration of \cite{Popper:80}, and are $T_{\rm eff} = 5920$\,K for
the primary and 3900\,K for the secondary, with estimated
uncertainties of 100\,K. The primary value is consistent with our
earlier spectroscopic estimates (Table~\ref{tab:teff}).

The distance to \vstar\ is listed also in Table~\ref{tab:absdim}, along
with other derived properties; it relies on an average out-of-eclipse
brightness of $V = 9.886 \pm 0.004$ based on the literature sources
cited above, corrected for extinction using $A(V) = 3.1
E(B-V)$. Separate distance calculations for the two components yield
consistent results.

%%%%%%%%%%%%%%%%%%%%%%%%%%%%%%%%%%%%%%%%%%%%%%%%%%%%%%%%%%%%%%%%%%%%%%%%%%%
\section{Comparison with theory}
\label{sec:models}
%%%%%%%%%%%%%%%%%%%%%%%%%%%%%%%%%%%%%%%%%%%%%%%%%%%%%%%%%%%%%%%%%%%%%%%%%%%

%%%%%%%%%%%%%%%%%%%%%%%%%%%%%%%%%%%%%%%%%%%%%%%%%%%%%%%%%%%%%%%%%%%%%%%%%%%
\subsection{Standard models}
\label{sec:models_standard}
%%%%%%%%%%%%%%%%%%%%%%%%%%%%%%%%%%%%%%%%%%%%%%%%%%%%%%%%%%%%%%%%%%%%%%%%%%%

\begin{figure}
\epsscale{1.15}
\plotone{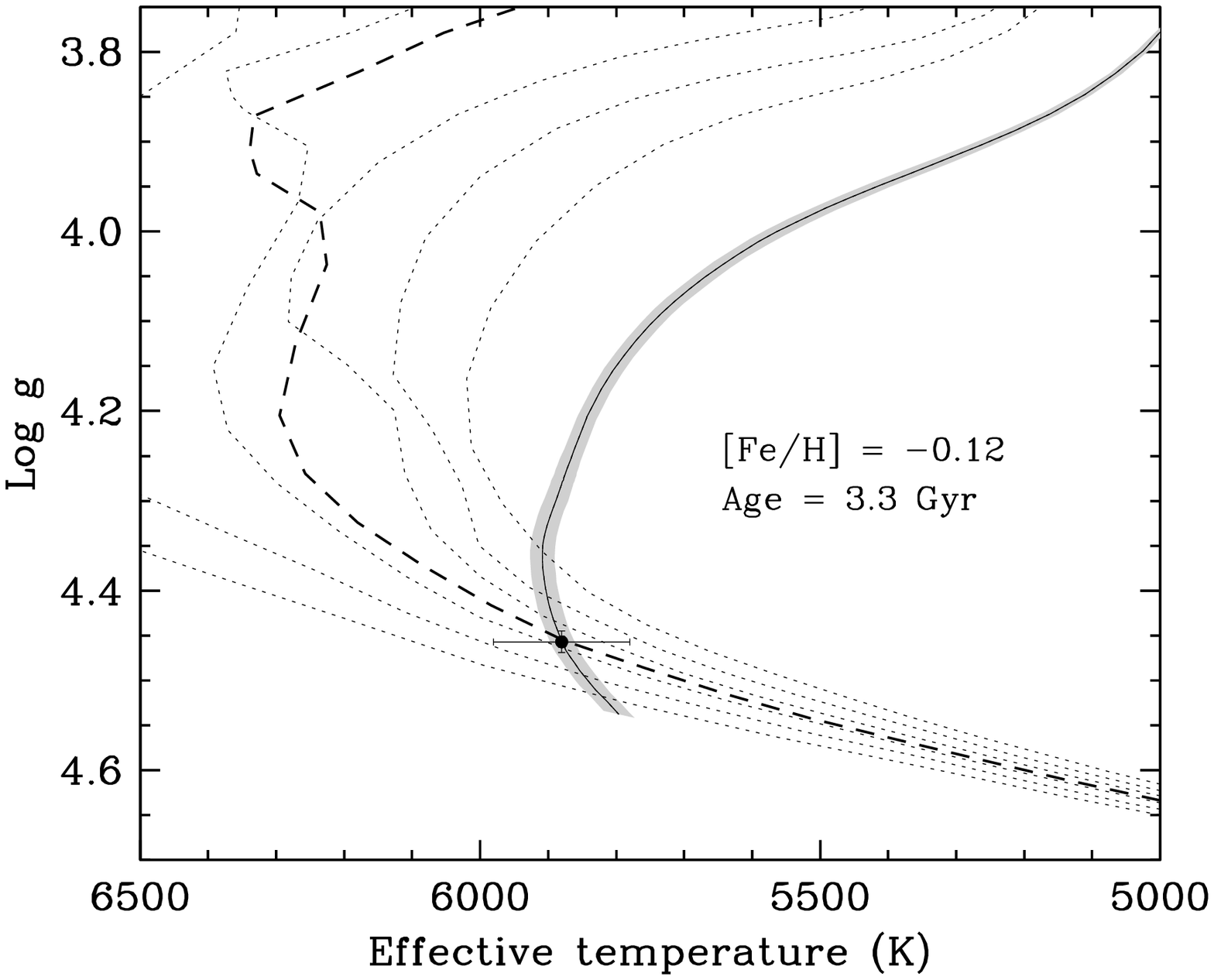}

\figcaption{Measurements for the primary of \vstar\ compared against
  models from the Yonsei-Yale series by \cite{Yi:01, Demarque:04} for the measured
  metallicity of ${\rm [Fe/H]} = -0.12$. The solid line is an
  evolutionary track for the measured mass, and the shaded area around
  it represents the uncertainty in the location of the track coming
  from the mass uncertainty. Isochrones from 1 to 6\,Gyr are shown
  with dotted lines, and the one rendered with a dashed line
  corresponds to the best fit for an age of about
  3.3\,Gyr.\label{fig:yale}}

\end{figure}

Our knowledge of the metallicity of \vstar\ presents an opportunity for
a stringent test of models of stellar evolution against our highly
accurate mass, radius, and temperature measurements, with one less
free parameter than is common in these types of comparisons. This is
particularly important in this case because the system contains an M
star, for which abundance analyses are usually very challenging and
generally unavailable.  A first test is shown in
Figure~\ref{fig:yale}, using the models from the Yonsei-Yale series
\citep{Yi:01, Demarque:04}. These models are intended for solar-type stars, and
adopt gray boundary conditions between the interior and the
photosphere that are adequate for stars more massive than about
0.7\,$M_{\odot}$, but become less realistic for lower-mass stars such
as the secondary of \vstar. Consequently, we compare them only against
the primary, which is very similar to the Sun. As shown in the figure,
an evolutionary track for the measured mass of the star and its
measured metallicity is in near perfect agreement with its temperature
and surface gravity, at an age of about 3.3\,Gyr. The star is
approaching the half-way point of its main-sequence phase. Consistent
with this old age, there is no sign of the \ion{Li}{1} $\lambda$6708
absorption line in the disentangled spectra of either star.

\begin{figure}
\epsscale{1.15}
\plotone{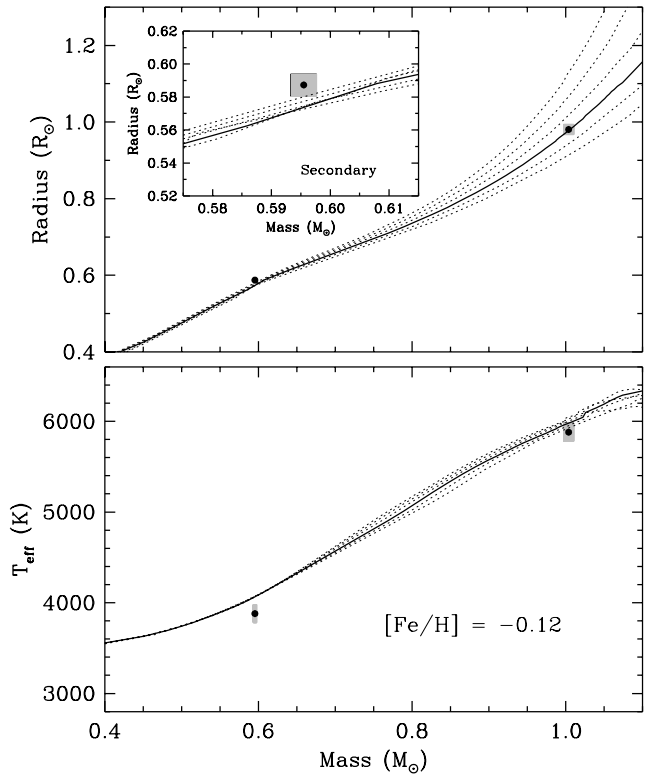}

\figcaption{Measured properties of \vstar\ compared with the Dartmouth
  models by \cite{Dotter:08}. \emph{Top:} Mass-radius diagram showing
  isochrones from 1 to 6\,Gyr for the measured metallicity of ${\rm
    [Fe/H]} = -0.12$, with the solid line representing the isochrone
  that best fits the primary star (3\,Gyr). The inset shows an
  enlargement around the secondary, which is seen to be larger than
  predicted. \emph{Bottom:} Mass-temperature diagram with the same
  isochrones as above.\label{fig:dartmouth}}

\end{figure}

Figure~\ref{fig:dartmouth} shows a comparison with model isochrones
from the Dartmouth series \citep{Dotter:08}, which are appropriate
both for solar-type and lower-mass stars. A 3\,Gyr isochrone computed
for the metallicity of the system reproduces the radius of the primary
star at its measured mass, but underestimates the size of the
secondary by about 2.5\% (see inset in the top panel of the
figure). This same isochrone is consistent with the temperature of the
primary, within its uncertainty, but slightly overestimates that of
the secondary. Similar anomalies in radius and temperature have been
seen in many other M dwarfs, and are attributed to the effects of
stellar activity and/or magnetic fields \citep[for a recent review of
  this phenomenon see][and references therein]{Torres:13}. One such
system of M dwarfs is YY\,Gem \citep{TorresRibas:02, Torres:10}, whose
two identical components happen to have virtually the same mass and
$T_{\rm eff}$ as the secondary of \vstar, but a radius that is 5\%
larger. While age and composition differences may be part of the
explanation, variances in the activity levels (YY\,Gem being much more
active) are likely to play a significant role as well.

\begin{figure}
\epsscale{1.05}
\plotone{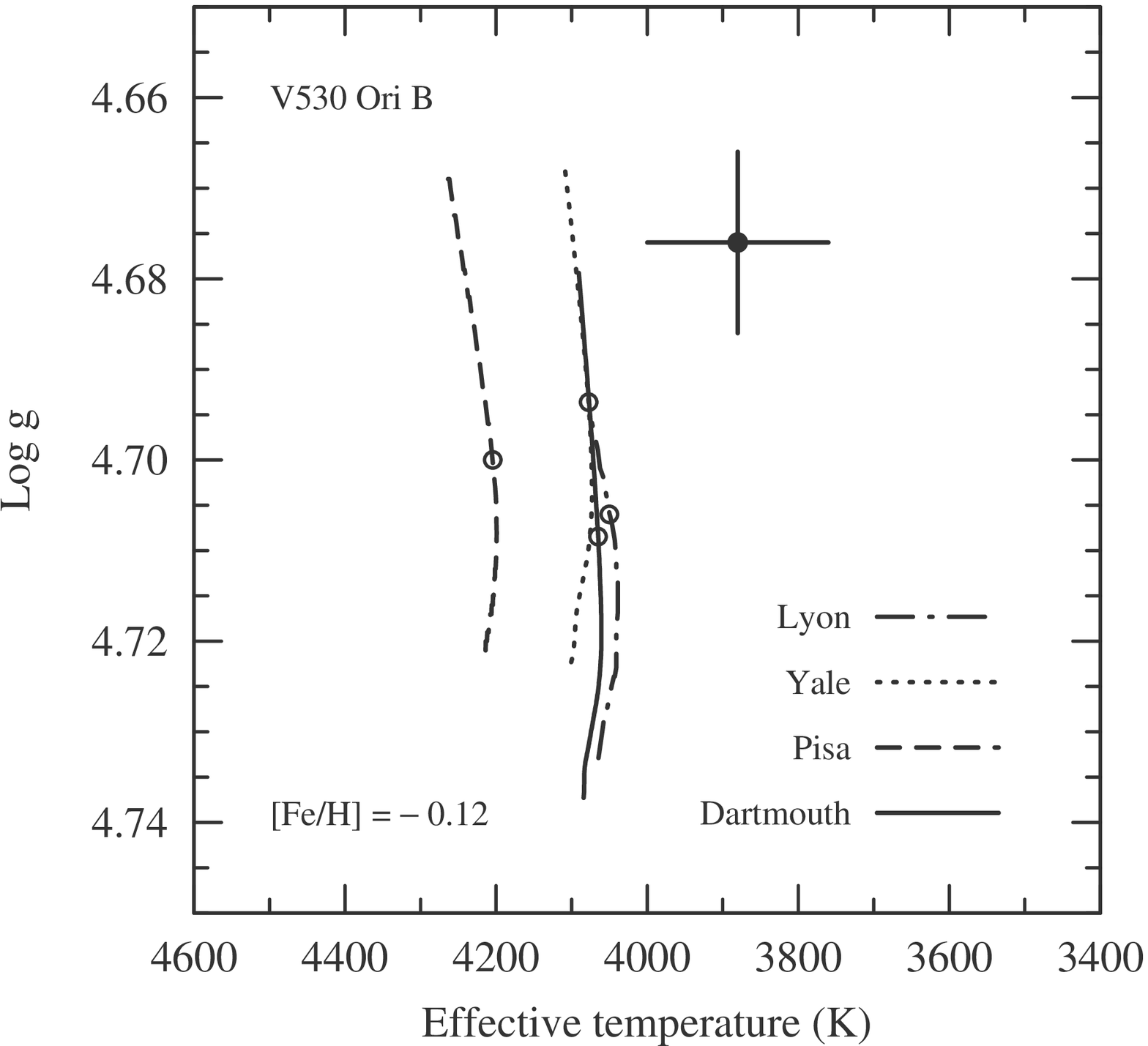}

\figcaption{Properties for the low-mass secondary of \vstar\ (solid
  circle with error bars) shown against evolutionary tracks for a mass
  of 0.6\,$M_{\sun}$ similar to that measured for the star, and ages
  of 140\,Myr to 10\,Gyr. Models represented are those from Lyon
  \citep{Baraffe:97, Baraffe:98}, Yale \citep{Spada:13}, and Pisa
  \citep{DellOmodarme:12}, interpolated to the measured metallicity of
  the system or at the nearest composition available (see text). Also
  shown for reference is a track from the Dartmouth series. Open
  circles on each track mark the properties of the secondary at the
  age predicted by models of the primary. In all cases the models
  underestimate the secondary radius (i.e., they overestimate $\log
  g$) and predict temperatures that are too
  hot.\label{fig:models_sec}}

\end{figure}

Several other series of models have been published in recent years
that incorporate realistic physical ingredients appropriate for
low-mass stars such as the secondary of \vstar\ (non-gray boundary
conditions, improved high-density/low-temperature equations of
state). These include the PARSEC models from the Padova series
\citep{Chen:14}, calculations from the Yale group \citep{Spada:13},
and from the Pisa group \citep{DellOmodarme:12}. Older models that are
also appropriate and are still widely used are those from the Lyon
group \citep{Baraffe:97, Baraffe:98}. Figure~\ref{fig:models_sec}
presents a comparison in the $\log g$ vs.\ $T_{\rm eff}$ diagram of
the measured properties for \vstar\,B against evolutionary tracks from
most of the above models for a mass of 0.6\,$M_{\sun}$, conveniently
very close to the measured mass of 0.5955\,$M_{\sun}$.  Tracks are
shown for ages from 140\,Myr to 10\,Gyr, with open circles marking the
predicted properties of the secondary at the best-fit age for the
primary in each model. We include also a 0.5955\,$M_{\sun}$ model from
the Dartmouth series, for reference. We point out, however, that such
comparisons are not always straightforward, or even possible in some
cases, due to coarseness of the model grids, limitations in the set of
parameters available (metallicity, mixing length parameter), and the
need to interpolate among existing models, which most likely limits
the accuracy. In particular, we have not compared against the Padova
models as only isochrones (but not yet evolutionary tracks) are
available. The Pisa track shown in Figure~\ref{fig:models_sec} is for
the highest metallicity available ($Z = 0.01$), which is marginally
lower than we measure for \vstar. For the Lyon models interpolation to
the measured metallicity of ${\rm[Fe/H]} = -0.12$ is only possible for
a mixing length parameter of $\alpha_{\rm ML} = 1.0$, whereas all
other models adopt a solar-calibrated value of $\alpha_{\rm
  ML}$. Additionally, there are differences in the interior
compositions adopted in all these calculations, and in many other
details that may explain why the predictions differ from model to
model, though a thorough discussion of these issues is beyond the
scope of this paper. Nevertheless, a common pattern seen in the figure
is that all models overestimate the temperature of the secondary star
by 4--8\%, and also overestimate its surface gravity, which means they
underestimate the radius (by about 2 to 4\%). These discrepancies are
in the same direction as found previously for many other low-mass
stars.

\begin{figure}
\epsscale{1.15}
\plotone{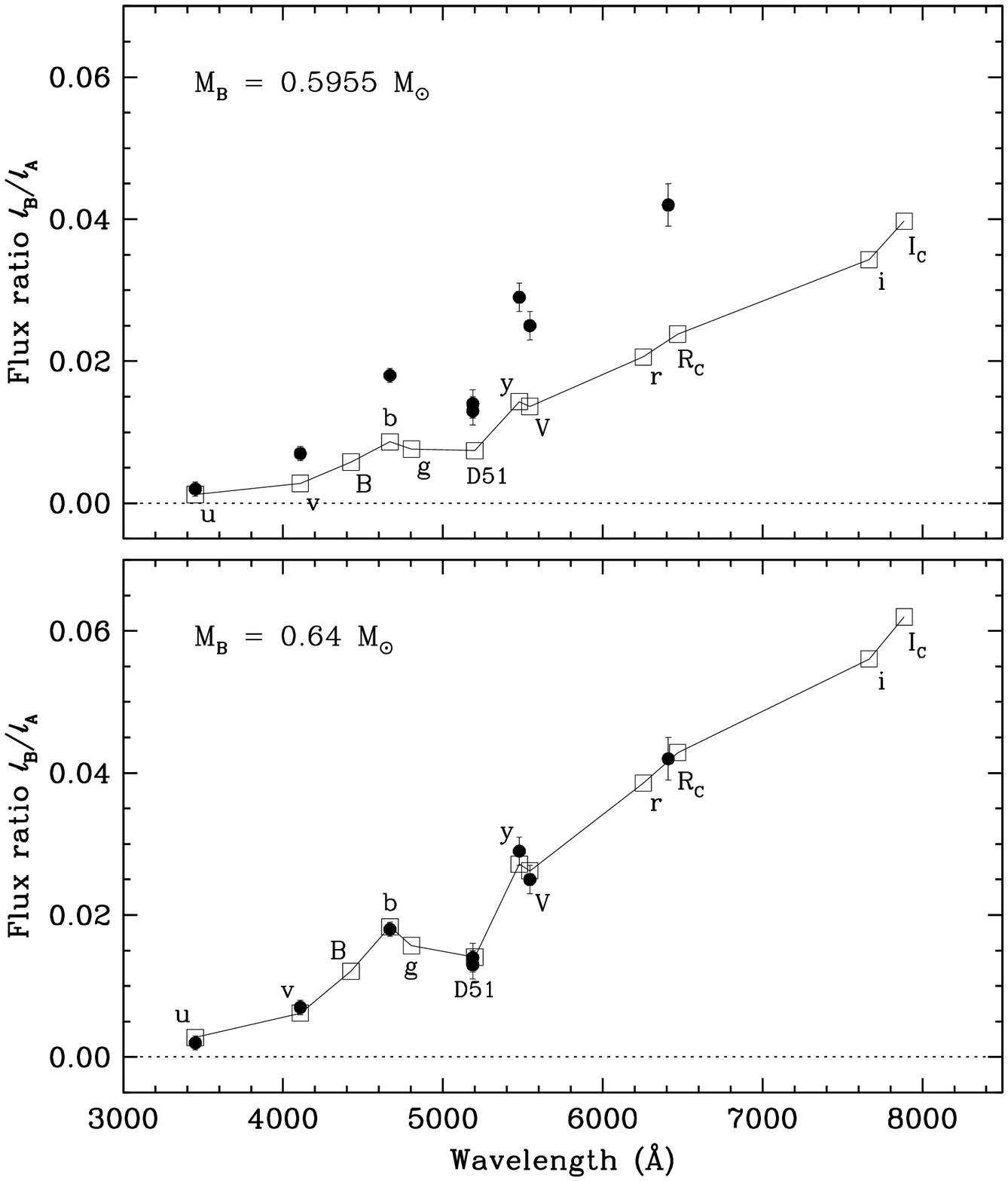}

\figcaption{\emph{Top:} Measured flux ratios ($\ell_{\rm B}/\ell_{\rm
    A}$) from our spectroscopic and photometric analyses of
  \vstar\ compared against predictions from the Dartmouth models of
  \cite{Dotter:08} for the measured masses of the two stars.
  Theoretical values for a number of standard photometric passbands
  are marked with open squares and connected with a line, and were
  computed using the same 3\,Gyr, ${\rm [Fe/H]} = -0.12$ best-fit
  isochrone from Fig.~\ref{fig:dartmouth}. Passbands labeled include
  Str\"omgren $uvby$, Sloan $gri$, Johnson-Cousins $BV(RI)_{\rm C}$,
  and D51, whose central wavelength coincides with the \ion{Mg}{1}\,b
  triplet, and therefore closely matches the spectroscopic window of
  our DS and TRES observations. \emph{Bottom:} Same as above, changing
  the secondary mass to be 0.64\,$M_{\sun}$ instead of the measured
  value of 0.5955\,$M_{\sun}$.\label{fig:lightratios}}

\end{figure}

Additional differences between models and observations for \vstar\ are
seen when comparing the secondary/primary flux ratios we estimated
spectroscopically and photometrically (Sect.~\ref{sec:spectroscopy}
and Sect.~\ref{sec:lightcurves}) against predictions for stars with
the exact masses we measure. We illustrate this in
Figure~\ref{fig:lightratios}, in which the predictions in several
standard photometric passbands are based on the same 3\,Gyr Dartmouth
isochrone that provided the best fit to the mass and radius of the
primary in Figure~\ref{fig:dartmouth}. Models systematically
underestimate all of the measured flux ratios by roughly a factor of
two, with the absolute deviations increasing toward longer
wavelengths. This is not entirely unexpected, given that the models
also fail to match the radius and temperature of the secondary star,
as well as its bolometric luminosity, which is overestimated.
Interestingly, we find that arbitrarily increasing the secondary mass
to $M_{\rm B} = 0.64$\,$M_{\sun}$ leads to predictions that agree
nearly perfectly with all of the measured flux ratios (bottom panel of
Figure~\ref{fig:lightratios}), from Str\"omgren $u$ to the value
measured from our KPNO spectra at $\sim$6410\,\AA, close to the
$R_{\rm C}$ band. This is unlikely to be a coincidence. We note,
though, that a mass for the secondary of $0.64$\,$M_{\sun}$ (nearly
7\% larger than measured, or $\sim$18$\sigma$) is implausibly large
given our observational uncertainties, and would not make the fit to
the other global properties ($R$, $T_{\rm eff}$) any better. The
reason for the underpredicted $\ell_{\rm B}/\ell_{\rm A}$ values may
be related to deficiencies in the temperature-color transformations
adopted in the Dartmouth models, which are based on PHOENIX model
atmospheres \citep{Hauschildt:99a, Hauschildt:99b}, and which are
known to degrade rapidly at optical wavelengths for cooler stars. Even
so, one might expect the predictive power of these models to be better
when considering flux ratio \emph{differences} between one wavelength
and another (e.g., the difference between $[\ell_{\rm B}/\ell_{\rm
    A}]_y$ and $[\ell_{\rm B}/\ell_{\rm A}]_{\rm D51}$), because those
rely on theory only in a differential sense. This is indeed what we
see in Figure~\ref{fig:lightratios}, and we take this to represent
indirect support for the accuracy of our light curve solutions in
Sect.~\ref{sec:lightcurves} (performed independently in each
passband), and therefore of the accuracy of the measured stellar
radii.

%%%%%%%%%%%%%%%%%%%%%%%%%%%%%%%%%%%%%%%%%%%%%%%%%%%%%%%%%%%%%%%%%%%%%%%%%%%
\subsection{Magnetic models}
\label{sec:models_magnetic}
%%%%%%%%%%%%%%%%%%%%%%%%%%%%%%%%%%%%%%%%%%%%%%%%%%%%%%%%%%%%%%%%%%%%%%%%%%%

A series of stellar models were computed using the magnetic Dartmouth
stellar evolution code \citep{Feiden:12, Feiden:13} to test the idea
that magnetic fields are responsible for the observed anomalies
between the secondary in \vstar\ and stellar models. The aim of the
present analysis is to first determine whether magnetic models are
able to provide a consistent solution for the two components of \vstar,
and then, if a consistent solution is identified, to establish whether
the conditions presented by the models are physically plausible.

\begin{figure}
\epsscale{1.15}
\plotone{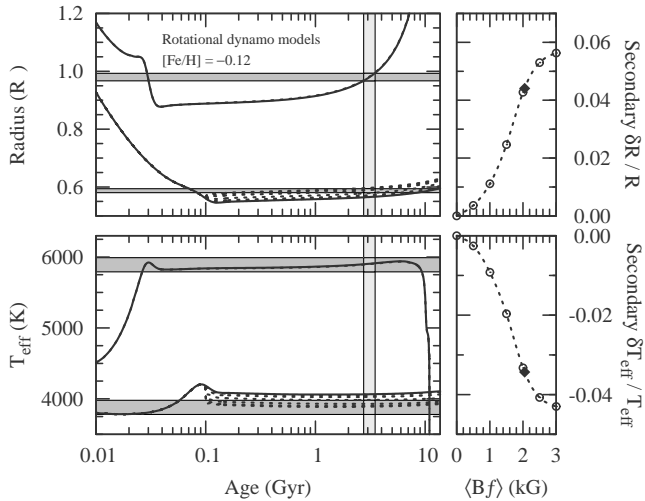}

\figcaption{\emph{Left:} Dartmouth models for the metallicity of
  \vstar\ compared against the measured radii and temperatures of the
  components, represented by the horizontal bands. Standard
  (non-magnetic) evolutionary tracks for the precise masses of the
  stars are drawn with solid lines, and models incorporating magnetic
  fields with a rotational dynamo prescription are drawn with dotted
  lines.  Field strengths for the secondary are $\langle Bf\rangle =$
  0.5, 1.0, 1.5, 2.0, 2.5, and 3.0 kG, and result in increasing
  departures from the standard models. A magnetic model with a field
  strength of 170 G is shown for the primary, but is nearly
  indistinguishable from the corresponding standard model. The best
  fit age range is shown by the vertical band.  \emph{Right:} Relative
  changes in radius ($\delta R/R$) and effective temperature ($\delta
  T_{\rm eff}/T_{\rm eff}$) for the secondary as a function of the
  strength of the magnetic field (see text). The best fit value is
  marked with a filled diamond. \label{fig:mag_rot}}

\end{figure}

Prior to implementing magnetic fields in the stellar evolution
calculations, as a check we re-assessed the performance of the
standard (i.e., non-magnetic) models from the magnetic Dartmouth code
owing to small differences with the original Dartmouth models of
\cite{Dotter:08}.  Comparisons were carried out in the age-radius and
age-$T_{\rm eff}$ planes for mass tracks computed at the precise
masses and metallicity of the \vstar\ stars. Figure~\ref{fig:mag_rot}
shows that properties of the primary star are well reproduced by the
model (represented with a solid line) between 2.7 and 3.5 Gyr,
yielding an age of $3.1 \pm 0.4$ Gyr, similar to our earlier finding.
As discussed before, the properties of the secondary are not
reproduced by the corresponding standard model.  Instead, theory
predicts a radius that is 3.7\% too small and a temperature that is
4.8\% too hot compared to observations.  Given that standard models
match the properties of the primary to a large degree, we began our
magnetic model analysis by assuming only the secondary is affected by
the presence of a magnetic field.

\begin{figure}
\epsscale{1.15}
\plotone{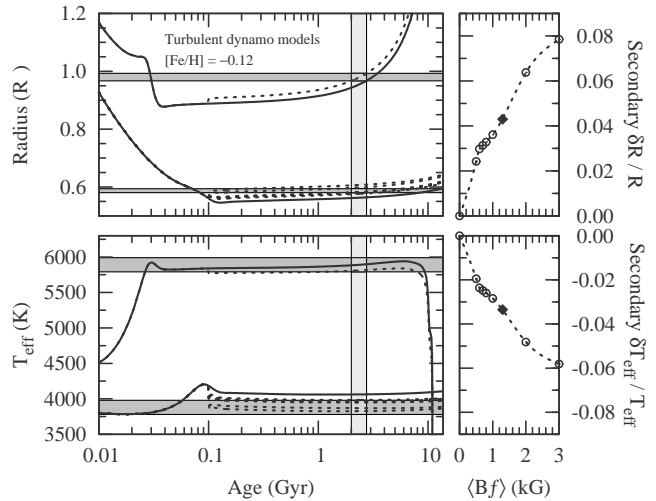}

\figcaption{Similar to Fig.~\ref{fig:mag_rot} but for magnetic models
  with a turbulent dynamo. Field strengths shown for the secondary are
  0.5, 0.6, 0.7, 0.8, 1.0, 2.0, and 3.0 kG (dotted lines). The
  magnetic model for the primary has $\langle Bf\rangle = 170$ G, and
  produces more noticeable changes in the radius and temperature of
  the star than the rotational dynamo model shown in
  Fig.~\ref{fig:mag_rot}.\label{fig:mag_turb}}

\end{figure}

A small grid of magnetic stellar models was computed at a fixed mass
(0.596\,$M_{\sun}$) and metallicity (${\rm [Fe/H]} = -0.12$) for
\vstar\,B.  Two procedures were used for modeling the influence of the
magnetic field on convection that are described by \cite{Feiden:13}.
These two procedures were designed to roughly mimic the effects of two
different dynamo actions: a rotational or shell dynamo
($\alpha$--$\Omega$) and a turbulent or distributed dynamo
($\alpha^2$).  All models utilized a dipole radial profile as the
influence of the magnetic field is only weakly dependent on the choice
of radial profile for stars with a radiative core and convective
envelope \citep{Feiden:13}.  For models using the rotational dynamo
procedure, values of the average surface magnetic fields were $\langle
Bf \rangle = $ 0.5, 1.0, 1.5, 2.0, 2.5, and 3.0 kG, while for the
turbulent dynamo the values were $\langle Bf \rangle =$ 0.5, 0.6, 0.7,
0.8, 1.0, 2.0, and 3.0 kG, in which $B$ is the photospheric magnetic
field strength and $f$ the filling factor. Corresponding mass tracks
are show with dotted lines in Figures~\ref{fig:mag_rot} and
\ref{fig:mag_turb}, with the relative changes in radius ($\delta R/R$)
and temperature ($\delta T_{\rm eff}/T_{\rm eff}$) of the secondary
indicated on the right as a function of the strength of the magnetic
field.

Results show that magnetic models of \vstar\,B can be made to reproduce
the observed properties assuming either dynamo procedure, with the
rotational dynamo suggesting $\langle Bf \rangle_{\rm B} = 2.1 \pm
0.4$~kG and the turbulent dynamo giving $\langle Bf \rangle_{\rm B} =
1.3 \pm 0.4$~kG. These values were calculated by extracting the
properties of each magnetic model computed at an age of 3.1 Gyr, and
generating curves using a cubic spline interpolation that give the
model radius and model temperature difference between the primary and
secondary as functions of $\langle Bf \rangle$ (right panels of
Figures~\ref{fig:mag_rot} and \ref{fig:mag_turb}). The spacing of the
magnetic field strength was 0.05 kG along the interpolated curves. We
then computed the $\chi^2$ value,
\begin{equation}
\nonumber
	\chi^2 = \left(\frac{R_{\rm obs} - R_{\rm mod}}{\sigma_R}\right)^2 +
	\left(\frac{\Delta T_{\rm eff,\, obs} - \Delta T_{\rm eff,\, mod}}{\sigma_{\Delta T}}\right)^2,
\end{equation}
at each point along the interpolated curve and took the resulting
minimum as the best-fit $\langle Bf \rangle$. For completeness, we
note that the minimum $\chi^2$ value we found is $\chi^2_{\rm min} =
0.4$.  Approximate errors for the permitted model $\langle Bf \rangle$
were determined by satisfying the condition $\chi^2 (\langle Bf
\rangle) = \chi_{\rm min}^2 + 1$.

As shown earlier, the primary star is active as well and may be
similarly influenced by its magnetic field, even though standard
models seem to be able to match the observed properties without that
effect. To test this, we generated magnetic models for the primary
star guided by an estimate of the field strength, described in the
next section, of $\langle Bf \rangle_{\rm A} = 170$~G. Results using
the rotational dynamo formulation are shown in
Figure~\ref{fig:mag_rot}, but produce only a negligible departure from
the standard model mass track. Figure~\ref{fig:mag_turb}, on the other
hand, demonstrates that the turbulent dynamo model causes a greater
level of radius inflation and temperature suppression in the primary.
Temperature suppression is such that agreement is nearly lost between
the model and the observations.  The age prediction is reduced to $2.4
\pm 0.4$ Gyr, and magnetic models of the secondary require moderately
stronger $\langle Bf \rangle$ values with the turbulent dynamo than in
the previous case. Performing the same procedure as before to generate
the best fit value, we obtained $\langle Bf \rangle_{\rm B} = 1.7 \pm
0.3$ kG.  However, in this case we found $\chi^2_{\rm min} = 3.5$,
indicating the final fit is poor. This is driven by the fact that the
temperature difference is more difficult to fit given the
significantly lower temperature of the primary model with a magnetic
field.

%%%%%%%%%%%%%%%%%%%%%%%%%%%%%%%%%%%%%%%%%%%%%%%%%%%%%%%%%%%%%%%%%%%%%%%%%%%
\subsubsection{Magnetic field strengths: empirical estimates}
\label{sec:activity}
%%%%%%%%%%%%%%%%%%%%%%%%%%%%%%%%%%%%%%%%%%%%%%%%%%%%%%%%%%%%%%%%%%%%%%%%%%%

Observational evidence for activity in \vstar\ is clear in the case of
the primary, and although no direct signs of it are seen for the very
faint secondary, we expect that star to be active as well. Approximate
magnetic field strengths for both stars were estimated as follows.
\cite{Saar:01} has shown there is a power-law relationship between
$\langle Bf \rangle$ and the Rossby number, $Ro \equiv P_{\rm
  rot}/\tau_{\rm c}$, where $P_{\rm rot}$ is the rotation period of
the star and $\tau_{\rm c}$ the convective turnover time. The Rossby
number for the primary may be estimated by noting that our
spectroscopic $v \sin i$ measurement suggests it is rotating either
synchronously or pseudo-synchronously.  We will assume the latter
here, although the difference is very small (see
Table~\ref{tab:absdim}). This leads to a rotation period of $P_{\rm
  rot} \approx 5.84$\,days based on the measured orbital eccentricity
\citep[see][]{Hut:81}. For $\tau_{\rm c}$ we must rely on theory.
Since the calibration of \cite{Saar:01} used convective turnover times
taken from the work of \cite{Gilliland:86}, we have done the same here
for consistency, and adopted (based on the temperature of 5890\,K)
$\tau_{\rm c} = 13.8 \pm 2$ days, with a conservative uncertainty. The
resulting Rossby number for \vstar\,A is $Ro = 0.423 \pm 0.067$.  A
similar calculation for the secondary gives $Ro = 0.116 \pm 0.005$
based on $\tau_{\rm c} = 50.3 \pm 2$ days \citep{Gilliland:86}, from
its temperature of 3880\,K, and assuming pseudo-synchronous rotation
(justified in view of the very short timescale for synchronization
compared to the age of the system; see
Sect.~\ref{sec:spectroscopy}). The \cite{Saar:01} relation then
projects a magnetic field strength for the primary of $\langle
Bf\rangle_{\rm A} = 170 \pm 140$\,G, and a value for the secondary of
$\langle Bf\rangle_{\rm B} = 830 \pm 650$\,G, where the uncertainties
account for all observational errors as well as the scatter of the
calibration.  The field strength for the secondary is not far from the
values required by the models in the previous section, suggesting the
theoretical predictions are at least plausible.

A consistency check on the empirically estimated $\langle Bf \rangle$
values may be obtained by relating these field strengths to X-ray
luminosities, and comparing them against a measure of the total X-ray
emission from \vstar\ detected by the ROSAT satellite.  Indeed,
\cite{Pevtsov:03} showed in a study of magnetic field observations of
the Sun and active stars that there is a fairly tight power-law
relationship between the X-ray luminosity and the total unsigned
surface magnetic flux, $\Phi = 4\pi R^2 \langle Bf\rangle$, which is
valid over many orders of magnitude. An updated relation restricted to
dwarf stars was presented by \cite{Feiden:13}. Using this latter
relation along with the measured stellar radii we obtain $\log L_{\rm
  X,A} = 28.63 \pm 0.59$ and $\log L_{\rm X,B} = 29.14 \pm 0.57$ (with
$L_{\rm X}$ in erg~s$^{-1}$). The sum of the X-ray luminosities
corresponds to $\log L_{\rm X,A+B} = 29.26 \pm 0.46$. The entry for
\vstar\ in the ROSAT All-Sky Survey Faint Source Catalog
\citep{Voges:00} lists a count rate of $0.0151 \pm
0.0072$\,cts\,s$^{-1}$ (0.1--2.4\,keV) and a hardness ratio of ${\rm
  HR1} = -0.43 \pm 0.37$ for the system, from a 465\,s exposure. The
corresponding total X-ray luminosity computed using the energy
conversion factor given by \cite{Fleming:95} and the distance in
Table~\ref{tab:absdim} is $\log L_{\rm X}{\rm (ROSAT)} = 29.06 \pm
0.33$. The good agreement between this measurement and the sum of the
individual X-ray luminosities, $\log L_{\rm X,A+B}$, may be taken as
an indication of the accuracy of the $\langle Bf \rangle$ values
reported above, even though their formal errors are large.

%%%%%%%%%%%%%%%%%%%%%%%%%%%%%%%%%%%%%%%%%%%%%%%%%%%%%%%%%%%%%%%%%%%%%%%%%%%
\section{Discussion}
\label{sec:discussion}
%%%%%%%%%%%%%%%%%%%%%%%%%%%%%%%%%%%%%%%%%%%%%%%%%%%%%%%%%%%%%%%%%%%%%%%%%%%

To the extent that our empirical magnetic field estimates above
represent the actual surface field strengths of the stars in \vstar,
it seems natural to require the models for \emph{both} components to
account for these effects. However, the way in which the influence of
magnetic fields on the stellar properties is treated in the models
seems to make a significant difference, particularly for the primary
star, and it is not at all clear which formulation is more
realistic. Given that this issue is at the heart of the long-standing
problem of radius inflation and temperature suppression in cool stars,
a careful consideration of the physical assumptions is in order.

Based strictly on the agreement with our empirical estimates, a
scenario whereby the primary star's magnetic field is generated by a
``rotational'' dynamo and the secondary by a more ``turbulent'' dynamo
would seem to be preferred.  In this case, the magnetic field of the
primary draws its energy largely from kinetic energy of (differential)
rotation, with the magnetic field rooted in a strong shear layer below
the convection zone (i.e., the tachocline), analogous to the mechanism
believed to drive the solar dynamo \citep{Parker:93,
  Charbonneau:97}. Convection is then inhibited by the stabilizing
effect that a (vertical) magnetic field has on a fluid
\citep{Gough:66, Lydon:95}. Given the similarity of \vstar\,A to the
Sun, the adoption of this magneto-convection formulation seems
justified. With a surface magnetic field strength $\langle
Bf\rangle_{\rm A} = 170$~G, the influence of a magnetic field on the
flow of convection is minimal and the structure of the model is
unaffected (see Figure~\ref{fig:mag_rot}), so that the magnetic model
produces results consistent with the non-magnetic model.

Concerning the secondary, both magnetic field formulations yield
agreement with the stellar properties ($T_{\rm eff}$ and $R$) at an
age defined by the properties of the primary (assuming the discussion
above holds). At face value the turbulent dynamo approach requires a
field strength ($\langle Bf\rangle = 1.3 \pm 0.4$\,kG) that is closer
to the empirically estimated value of $\langle Bf\rangle_{\rm B} =
0.83 \pm 0.65$\,kG than the alternate approach with a rotational
dynamo (which predicts $\langle Bf\rangle = 2.1 \pm 0.4$\,kG). The
accuracy of the empirical value is difficult to assess and depends
strongly on the reliability of the \cite{Saar:01} calibration. The
turbulent dynamo formulation simplistically assumes that the energy
for the magnetic field is provided by kinetic energy available in the
larger scale convective flow. Convection is then made less efficient
as energy is diverted away from convecting fluid elements thereby
impeding their velocity and thus reducing the total amount of
convective energy flux \citep[e.g.,][]{Durney:93, Chabrier:06,
  Browning:08}. Precisely how this conversion is achieved (e.g.,
through turbulence, helical convection, or feedback generated by the
Lorentz force) is not explicitly defined in the stellar models.

While consistency between the estimated surface magnetic field
strength and that required by the models is encouraging, it is not
clear that the dynamo mechanism at work in \vstar\,B should be any
different from that in \vstar\,A. Both stars possess a radiative core
and a convective outer envelope and thus, presumably, a stable
tachocline in which to produce a magnetic field through an interface
dynamo. Furthermore, the presence of a stable tachocline is not
necessarily a strict condition for a solar-like dynamo
\citep{Brown:10}. Therefore, there is no reason \emph{a priori} to
believe that the stars should have a different dynamo mechanism. If we
instead assume that the primary also has a dynamo driven by
convection, then the structural changes imparted by the magnetic field
become significant, even for a modest 170~G magnetic field at the
surface.  Changes induced on the primary are such that models of the
primary and secondary cannot be made to agree at the same age, leaving
us with precisely the same problem that we were looking to correct
with the magnetic models.

A possible reason to expect a different dynamo mechanism would be if
differential rotation were somehow suppressed in the secondary star.
Quenching of differential rotation has been observed in detailed
magneto-hydrodynamic simulations as a result of Maxwell stresses
produced by an induced magnetic field \citep{Browning:08}. On the
other hand, simulations of a Sun-like star with an angular velocity
similar to \vstar\,A do not demonstrate this quenching
\citep{Brown:10}, so we may posit that the primary star has a dynamo
driven by differential rotation, as we initially supposed. Although
the two components of \vstar\ are likely rotating with a similar
angular velocity, convective velocities in the secondary are slower,
leading to convective flows that are more susceptible to the influence
of the Coriolis force. This could then drive strong magnetic fields
that also quench the differential rotation. Unfortunately, assessing
the level of differential rotation on the secondary is not currently
possible.

\cite{Browning:08} predicts that when differential rotation is
quenched, the large scale axisymmetric component of the magnetic field
should account for a larger fraction of total magnetic energy. Using
the empirical scaling relations of \cite{Vidotto:14}, we estimated the
large scale magnetic field component on each star using our derived
X-ray luminosities.  We find that the large scale component of the
magnetic field (taken to be perpendicular to the line of sight) makes
up 6\% and 12\% of the total magnetic energy,
corresponding to $\langle Bf\rangle_{\perp} = 10$~G and 100~G for
\vstar\,A and B, respectively. While the trend is consistent with the
secondary having a more significant large scale field component (in
terms of total magnetic energy contribution), it is not possible to
say whether this is the result of different dynamo actions.

In summary, while many critical aspects of the problem are still not
understood, the arguments above seem to support a picture in which the
models are able to match the measured temperatures and radii of the
components with the magnetic field playing little role in changing the
structure of the primary star (i.e., consistent with it having a
rotational dynamo). The nature of the magnetic field on the secondary
is less clear, with the observations perhaps favoring a distributed
(turbulent) dynamo over a rotational one, but not at a very
significant level.

Other consequences of magnetic fields on structure of the stars in
\vstar\ appear small: the predicted apsidal motion constant corresponds
to an apsidal motion period of $U = 19,\!400$ yr for a magnetic
secondary (both dynamo types), not very different from the value of
$19,\!100$ yr computed with no magnetic fields. The observed value
from Sect.~\ref{sec:ephemeris} is unfortunately much too imprecise for
a meaningful comparison.  We note that the properties of the system
are such that the contribution to the apsidal motion from General
Relativity effects \citep[e.g.,][]{Gimenez:85} is expected to dominate
(72\%) over the classical terms from tidal and rotational distortion.

A larger effect of magnetic fields is seen on the convective turnover
time.  The Dartmouth models yield $\tau_c = 16$ days for the primary
star, somewhat longer than other estimates mentioned earlier, and
values for the secondary of 50.5 days (standard, non-magnetic), 49.3
days (rotational dynamo), and 65.4 days (turbulent dynamo).

%%%%%%%%%%%%%%%%%%%%%%%%%%%%%%%%%%%%%%%%%%%%%%%%%%%%%%%%%%%%%%%%%%%%%%%%%%%
\section{Concluding remarks}
\label{sec:conclusions}
%%%%%%%%%%%%%%%%%%%%%%%%%%%%%%%%%%%%%%%%%%%%%%%%%%%%%%%%%%%%%%%%%%%%%%%%%%%

With masses and radii determined to better than 0.7\% and 1.3\%,
respectively, and a secondary of spectral type M1, \vstar\ joins the
ranks of the small group of eclipsing binary systems containing at
least one low-mass main-sequence star with well-measured
properties. What distinguishes this example is that the chemical
composition is well known from our detailed analysis of the
disentangled spectrum of the primary component, which is an easily
studied G1 star.  Investigations of most other systems containing M
stars have struggled to infer metallicities directly from the
molecule-ridden spectra of the M stars, or by more indirect
means. Knowledge of the metallicity removes a free parameter in the
comparison with stellar evolution models that permits a more
meaningful test of theory, as we have done here. We have also made a
special effort to establish an accurate temperature for the primary
star by measuring it in several different ways, as the $T_{\rm eff}$
value for the secondary hinges on it, as does the entire comparison
with models.

Both the Yonsei-Yale and the Dartmouth models provide a good match to
the primary star at the measured metallicity, suggesting that both its
temperature and metallicity are accurate.  On the other hand, we find
that standard models from the Dartmouth series underpredict the radius
and overpredict the temperature of the secondary by several percent,
as has been found previously for many other cool main-sequence stars.
Magnetic models from the same series succeed in matching the observed
radii and temperatures of both stars at their measured masses with
surface magnetic fields for the secondary of about 1--2\,kG in
strength, fairly typical of early M dwarfs, and an age of some 3\,Gyr.
These field strengths are not far from what we estimate empirically
for \vstar\,B on the basis of the Rossby numbers. The agreement is
reassuring, and suggests that we are closer to understanding radius
inflation and temperature suppression for convective stars, not only
qualitatively but also quantitatively.  Earlier quantitative evidence
in this direction was presented by \cite{Feiden:12, Feiden:13,
  Feiden:14}, also for the Dartmouth models, with the present case
being perhaps a stronger test in that our estimates of the individual
magnetic field strengths used somewhat weaker assumptions. \vstar\ is
thus a key benchmark system for this sort of test.  Questions remain,
however, about the exact nature of the magnetic fields and how their
effect on the global properties of the stars should be treated in the
models (rotational dynamo, turbulent dynamo, or some other
prescription).

\acknowledgments

We are grateful to P.\ Berlind, M.\ Calkins, R.\ J.\ Davis,
G.\ Esquerdo, D.\ Latham, A.\ Milone, and R.\ Stefanik for help in
obtaining the CfA observations of \vstar\ with the DS and with TRES,
and to R.\ J.\ Davis and J.\ Mink for maintaining the CfA echelle data
bases over the years.  A.\ Bieryla is acknowledged for help with the
initial spectroscopic solutions using the TRES spectra. The anonymous
referee provided very helpful comments on the manuscript. The authors
also wish to thank Bill Neely, who operates and maintains the NFO
WebScope for the Consortium, and who handles preliminary processing of
the images and their distribution. We also thank G.\ Berard,
S.\ Bouley, M.\ Y.\ Bouzid, T.\ H.\ Dall, L.\ M.\ Freyhammer,
E.\ Johnsen, H.\ J{\o}rgensen, R.\ Leguet, C.\ Papadaki,
J.\ D.\ Pritchard, S.\ Regandell, and C.\ Sterken for their assistance
in gathering the photometric observations with the Str\"omgren
Automatic Telescope at ESO. JVC participated fully in the data
collection and analysis up to the time of his death, but bears no
responsibility for the final text of this paper. Finally, we also
thank Ian Czekala for computing the PHOENIX library of calculated
spectra used here. GT acknowledges partial support for this work from
NSF grant AST-1007992.

\end{document}